\pdfoutput=1
\documentclass[12pt]{article}

\usepackage[dvips]{graphicx}
\usepackage{amsmath}
\usepackage{amsfonts}
\usepackage{amssymb}
\usepackage{hyperref}
\usepackage{color}
\usepackage[small]{caption}
\usepackage{cite}

\numberwithin{equation}{section}

\newcommand{\bea}{\begin{eqnarray}}
\newcommand{\eea}{\end{eqnarray}}
\newcommand{\be}{\begin{equation}}
\newcommand{\ee}{\end{equation}}
\newcommand{\nn}{\nonumber}
\newcommand{\HH}{\mathcal H}

\begin{document}

\title{\vspace{-2cm} 
{\normalsize
\flushright CERN-PH-TH-2014-149 \\
\vspace{-.4cm}\flushright DESY 14-143 \\
\vspace{-.4cm}\flushright LAPTH-073/14\\}
\vspace{0.6cm} 
\bf  Structure formation with massive neutrinos: going beyond linear theory\\[8mm]}

\author{Diego Blas$^1$, Mathias Garny$^1$, Thomas Konstandin$^2$, Julien Lesgourgues$^{1,3,4}$\\[2mm]
{\normalsize\it  $^1$ CERN Theory division,}\\[-0.05cm]
{\normalsize\it  CH-1211 Geneva 23, Switzerland}\\[1.5mm]
{\normalsize\it  $^2$ Deutsches Elektronen-Synchrotron DESY,}\\[-0.05cm]
{\normalsize\it  Notkestra\ss{}e 85, 22603 Hamburg, Germany}\\[1.5mm]
{\normalsize\it $^3$Institut de Th\'eorie des Ph\'enom\`enes Physiques, 
 EPFL,}\\[-0.05cm]
{\it\normalsize CH-1015 Lausanne, Switzerland}\\
{\normalsize\it $^4$LAPTh (CNRS-Universit{\'e} de Savoie),}\\[-0.05cm]
{\it\normalsize B.P.\ 110, F-74941 Annecy-le-Vieux Cedex, France}
}

\maketitle

\begin{abstract}
\noindent
We compute non-linear corrections to the matter power spectrum taking the time- and scale-dependent free-streaming length of neutrinos into account. We adopt a hybrid scheme that matches the full Boltzmann hierarchy to an effective two-fluid description at an intermediate redshift. The non-linearities in the neutrino component are taken into account by using an extension of the time-flow framework. We point out that this remedies a spurious behaviour that occurs when neglecting non-linear terms for neutrinos. This behaviour is related to how efficiently short modes decouple from long modes and can be traced back to the violation of momentum conservation if neutrinos are treated linearly. Furthermore, we compare our results at next to leading order to  various other methods and quantify the accuracy of the fluid description. Due to the correct decoupling behaviour of short modes, the two-fluid scheme is  a suitable starting point to compute higher orders in perturbations or for resummation methods.
\end{abstract}

\newpage

\section{Introduction}

Future large-scale structure surveys are expected to return a wealth of information on the ingredients of the cosmological model describing our universe \cite{Amendola:2012ys,Abell:2009aa}. Remarkably, not only will they explore cosmological questions, such as the clustering properties of dark matter and dark energy or possible deviations from Einstein's gravity, but they may also enable us to determine the absolute neutrino mass scale~\cite{Lesgourgues:1519137}. In fact, the total neutrino mass $M_\nu \equiv \sum_i m_{\nu,i}$, with $m_{\nu,i}$ being the mass of each neutrino species, is known to be greater than $0.06$~eV given the neutrino oscillation data \cite{GonzalezGarcia:2012sz}.  The cosmological evolution is sensitive to  this quantity  and currently imposes the bound  $M_\nu <0.23$~eV (95\% CL) \cite{Ade:2013zuv}. The effects of a neutrino mass in this range of scales are very difficult to measure in the laboratory ($\beta$-decay) experiments~\cite{Osipowicz:2001sq}.  However, even the smallest possible mass has an impact on the total matter power spectrum by at least around 5\% at the scales of baryon acoustic oscillations (BAO) and redshift zero~\cite{Lesgourgues:1519137}. This minimum effect is expected to be detectable by the Euclid satellite at the level of $2-3$ $\sigma$~\cite{Audren:2012vy,Basse:2013zua}. Hence, it seems plausible that cosmological observations will be able to close the window in $M_\nu$ and make a major contribution to our understanding of the neutrino sector.

To achieve the previous goal (and to extract any other information from galaxy surveys with precision) there are many effects that need to be clarified. In particular, it will be necessary to understand with better precision than today several aspects of galaxy bias, redshift space distortions, and non-linear corrections to observables as the matter power spectrum or mass functions (see e.g. \cite{Weinberg:2012es,Saito:2014qha,Biagetti:2014pha,Baldauf:2014fza,Vlah:2013lia}). The impact of massive neutrinos in some of these effects (and possible degeneracies with other phenomenology) have been studied for example in \cite{Biagetti:2014pha,Marulli:2011he,Baldi:2013iza,LoVerde:2014rxa,LoVerde:2013lta,VillaescusaNavarro:2011ys,Dvorkin:2014lea,Rossi:2014wsa,Villaescusa-Navarro:2013pva,Castorina:2013wga,Costanzi:2013bha}. 

In this work we focus on the non-linear corrections to the matter power spectrum. Including neutrinos poses an additional challenge due to their large velocity dispersion. Several approximate schemes have been devised to take this hot component into account in $N$-body simulations  (see e.g. \cite{Bird:2011rb,Wagner:2012sw,Borde:2014xsa, Hannestad:2011td}). Those simulations require long computational times, and it would  be ideal to find faster ways to compute certain features of the non-linear matter power spectrum with massive neutrinos, with analytic or semi-analytic methods, hopefully on a time-scale of the order of a few seconds, suitable for a Monte Carlo Markov Chain exploration of the model parameter space. A first possibility in this direction is using cosmological perturbation theory since the effects of massive neutrinos are distinguishable at the scales and redshifts where this method has proven to be useful \cite{Bernardeau:2013oda}. 

Cosmological perturbation theory is based on the idea that the matter components of the universe can be treated as media with  originally small density contrasts $\delta_i$ that can be used as  small parameters in a perturbative expansion \cite{Bernardeau:2013oda,Bernardeau:2001qr}. The different $\delta_i$ have different amplitudes and evolution. The cold dark matter (CDM) component $\delta_{c}$ is the most important one during matter domination, and thus for the generation of the large scale structure of the universe. The first calculations including massive neutrinos were presented in Saito et al.~\cite{Saito:2008bp} and Wong~\cite{Wong:2008ws} where the density perturbation of neutrinos $\delta_\nu$ was kept at linear order while $\delta_{cb}$ (CDM + baryon density contrast) was treated non-linearly at the one-loop level. In Saito et al.~\cite{Saito:2008bp} the kernels used to propagate non-linear corrections were approximated  by the ones of pure Einstein de Sitter (EdS) cosmology, while Wong~\cite{Wong:2008ws} improved these kernels to include some of the massive neutrino effects. Subsequently, \cite{Lesgourgues:2009am,Upadhye:2013ndm} considered a variation of the time-flow equations \cite{Pietroni:2008jx}, taking the time- and scale-dependent impact of the linearized neutrino perturbations into account.

Given that the corrections from the neutrino component are already small at linear order, one may consider that this is a safe approximation and neglect higher order corrections. This is problematic for the following reason. A crucial property of non-linear corrections is that in the large-scale limit $k\ll k_{nl}$ (where $k_{nl}$ is a characteristic wavenumber at which non-linear corrections become of order one), these corrections are  suppressed by a factor $k^2/k_{nl}^2$ relative to the linear power spectrum. This screening of small-scales at large scales  follows from momentum conservation~\cite{Goroff:1986ep,Rees:1982,Peebles:1980}. By approximating the neutrino component by its linear contribution, momentum conservation is violated, and the sensitivity of long-modes to short-modes is parametrically enhanced. The consequence is a pathological behaviour in the small-$k$ limit which has a moderate impact on one-loop predictions, but would be catastrophic when pushing calculations to higher level. Thus, a treatment beyond the linear order of the neutrino component is necessary to give accurate predictions. This is challenging because of  the free-streaming of cosmic neutrinos, which renders a fluid description much less accurate than for the CDM component.

A multi-fluid  (CDM and neutrinos) description has been discussed in \cite{Shoji:2010hm,Lesgourgues:2011rh}, and  \cite{Shoji:2009gg} addressed the case of a second fluid with constant Jeans length (in contrast with the time-varying neutrino free-streaming length).  An alternative approach is to describe massive neutrinos as a superposition of several perfect fluids \cite{Dupuy:2013jaa}. The strategy in the present work to properly treat the non-linear corrections is to use the full Boltzmann hierarchy at early times, and to match to a two-fluid description at small redshift ($z\sim 25$).

In summary, our motivations are (i) to obtain a consistent computation of non-linear corrections to the power spectrum in presence of massive neutrinos, (ii) to cross-check existing approximation schemes of the time-evolution and quantify errors due to linear approximations for neutrinos, especially in view of the potentially spurious behaviour described above, and (iii) to develop a framework that provides a suitable basis for computing higher non-linear orders in the future. We present the set of two-fluid non-linear equations that we want to solve in section~\ref{sec:2}. In section~\ref{sec:3}, we discuss analytic arguments to explain why this approach is consistent, and free of unphysical divergence issues on large scales. In section \ref{sec:4}, we present numerical solutions and compare them with the results of other approaches. Section~\ref{sec:5} contains our conclusions. The Appendix contains an alternative formulation of the evolution equations.

\section{Two-fluid equations \label{sec:2}}

In this section we set up the fluid description for the neutrino component and the time-flow equations. 
As a first step, one has to establish to which extent a two-fluid scheme with an effective sound speed can approximate the full solution of the Boltzmann equation. 
This has been studied in detail before in \cite{Shoji:2010hm}, where the fluid description has been applied throughout the cosmological evolution. It was found that this scheme is accurate at the level of about $10\%$ for the neutrino density and velocity. Since we are interested in a higher precision, this may indicate that it is not appropriate to neglect higher moments of the neutrino distribution function. Actually, some more precise approximations involving one more moment have also been studied in details in \cite{Hu:1998kj,Shoji:2010hm,Lesgourgues:2011rh}. They amount in describing neutrinos as an imperfect fluid with an effective viscosity coefficient.

However, in our context, it is important to realise that non-linear effects become important only at low redshift $z \lesssim 10$, while the higher moments of the neutrino distribution are suppressed for $z< z_{nr}\sim 10^2$ by powers of $T_\nu/m_\nu$. Therefore, we use a hybrid scheme based on the full Boltzmann solution at high redshift, and on a two-perfect-fluid scheme that includes an effective pressure term for the neutrino component at low redshift. The matching can be done at some redshift in the range $10 \ll z_{match} \ll z_{nr}$. We used $z_{match}=25$. It turns out that this scheme is sufficiently accurate for our purposes (about $0.1\%(1\%)$ for the CDM($\nu$) component at $k=0.1\,h/$Mpc, see Sec.\,\ref{sec:4}). In the following we discuss the fluid scheme we use at small redshift for computing non-linear corrections.

\subsection{Two-fluid non-linear equations}

For each neutrino eigenstate $i$, the non-relativistic transition takes place when the mean neutrino energy becomes smaller than the
neutrino mass, at a redshift given by \cite{Shoji:2010hm,Lesgourgues:1519137}
\begin{equation}
 1+z_{nr,i} \simeq 1890 \frac{m_{\nu,i}}{1\text{eV}}\;.
\end{equation}
For $z\ll z_{nr,i}$, the fraction of the total matter energy density in the form of neutrinos becomes constant,
\begin{equation}
 f_\nu \equiv \frac{\Omega_\nu}{\Omega_m} = \frac{1}{\Omega_m^0 h^2} ~ \frac{\sum_i
m_{\nu,i}}{93.14~\mbox{eV}}\,,
\end{equation}
where $\Omega_m^0 \equiv \Omega_m(z=0)$ is the total matter density today in units of the critical density.
The total density contrast is given by
\begin{equation}
\label{eq:totaldelta}
\delta = f_\nu\delta_\nu + (1-f_\nu)\delta_{cb}\,,
\end{equation}
where $\delta_i=\delta\rho_i/\rho_i$, and $\delta_{cb}$ corresponds to the sum
of baryons and CDM.

We wish to treat neutrinos using perfect fluid equations (the accuracy of this approximation is discussed later, in Sec.~\ref{sec:4}). 
Hence, we must introduce an effective neutrino sound speed, even if $\frac{\delta p_i}{\delta \rho_i}$ depends on space coordinates for actual free-streaming neutrinos. Fortunately, our focus is on non-linear corrections at large scales and at
low redshifts $z\ll z_{nr}$. In this limit, there is a simple asymptotic relation between the squared sound speed $\frac{\delta p_i}{\delta \rho_i}(k,z)$, the velocity dispersion (or anisotropic stress/pressure) $\sigma_{\nu, i}(k,z)$, the equation of state parameter $w_i(z)$, the squared adiabatic sound speed $c_{g,i}^2(z) \equiv \frac{\dot{\bar{p}}_i(z)}{\dot{\bar{\rho}}_i(z)}$, and the temperature-to-mass ratio $\frac{T_\nu(z)}{m_{\nu,i}}$ (see e.g. \cite{Shoji:2010hm}),
\begin{equation}
\frac{\delta p_i}{\delta \rho_i} = \frac{5}{9}\sigma_{\nu,i}^2 = c_{g,i}^2 =  \frac{5}{3} w_i = \frac{5}{3}\frac{5\zeta(5)}{\zeta(3)}\left(\frac{T_\nu}{m_{\nu,i}}\right)^2  \;,
\qquad T_\nu \ll m_{\nu,i}~.
\end{equation}
We refer to the square root of this common limit simply as the neutrino sound speed, $c_{s,i}(z)\equiv 2.680 \frac{T_\nu}{m_{\nu,i}}$. It should not be confused with the root mean square of the neutrino particle velocity, $c_{\nu,i}(z) \equiv \langle p \rangle / m = 3.15 ~T_\nu/m_{\nu,i}$.
For convenience, the neutrino sound speed can be
expressed in terms of the free-streaming scale
\cite{Shoji:2010hm,Wong:2008ws}
\begin{equation}\label{kFS}
  k_{FS,i}(z) \equiv \sqrt{\frac{3\Omega_m}{2}}\frac{{\cal H}}{c_{s,i}(z)} \simeq
\frac{0.908}{(1+z)^\frac12}\frac{m_{\nu,i}}{1\,\mbox{eV}}\sqrt{\Omega_m^0}\, h /
\mbox{Mpc}\;.
\end{equation}
In principle the free-streaming scale is different for each neutrino species. For
simplicity, we consider the case of three degenerate species and drop the index $i$, although
all results can be generalized in a straightforward way to include several distinct
neutrino fluid components. 

The neutrinos interact gravitationally with the baryon-CDM component, which we describe by a 
pressureless perfect fluid. The Euler and continuity equations for the density contrast $\delta_i$ and
velocity divergence $\theta_i$ of the different species (neglecting vorticity) are
\begin{eqnarray}
  \dot\delta_{cb} + \theta_{cb} &=& -\alpha\theta_{cb}\delta_{cb} \,, \\
  \dot\theta_{cb} + {\cal H}\theta_{cb} + \frac32 {\cal H}^2\Omega_m [
f_\nu\delta_\nu + (1-f_\nu)\delta_{cb} ] &=& -\beta\theta_{cb}\theta_{cb}\,, \\
  \dot\delta_{\nu} + \theta_{\nu} &=& -\alpha\theta_{\nu}\delta_{\nu} \,, \\
  \dot\theta_{\nu} + {\cal H}\theta_{\nu} + \frac32 {\cal H}^2\Omega_m [
f_\nu\delta_\nu + (1-f_\nu)\delta_{cb} ] -k^2c_s(\tau)^2\delta_{\nu}&=&
-\beta\theta_{\nu}\theta_{\nu}\,,
\end{eqnarray}
where derivatives are w.r.t. the conformal time $d\tau=dt/a$, ${\cal H}=\dot
a/a$, and the right-hand side contains the usual non-linear terms as well as
convolution integrals in wavenumber space\footnote{We follow the conventions of
\cite{Bernardeau:2001qr} except for the normalization of Fourier integrals, where we use the integration measures $d^3q/(2\pi)^3$ and $d^3x$, respectively, as well as $\langle\delta_k\delta_{k'}\rangle=(2\pi)^3\delta(k+k')P(k)$.}
\begin{equation}
  \alpha\theta_{i}\delta_{i} \equiv \int \!\frac{d^3p}{(2\pi)^3} \frac{d^3q}{(2\pi)^3}\, (2\pi)^3\delta(k-p-q) \alpha(k,p,q) \theta_i(p,\eta)\delta_i(q,\eta) \,,
\end{equation}
with vertices
\begin{equation}
  \alpha(k,p,q) = \frac{k\cdot p}{p^2}, \qquad \beta(k,p,q) = \frac{k^2 p\cdot q}{2p^2q^2} \;.
\end{equation}

The Euler and continuity equations can be brought into the form
\begin{equation}
\label{eq:PTinvectors}
\frac{\partial}{\partial\eta}\psi_a + \Omega_{ab}(k,\eta)\psi_b =      
\gamma_{abc}\psi_b\psi_c \,,
\end{equation}
where $\eta\equiv\ln(a)$ (in the following we use the notation $f'\equiv d f/ d\eta$)
\begin{equation}
\begin{split}
 \psi_1 \equiv \delta_{cb}, \quad \psi_2 \equiv&  -\theta_{cb}/{\cal H}, \quad \psi_3 \equiv 
\delta_\nu, \quad \psi_4 \equiv -\theta_{\nu}/{\cal H} \;,\\
 \gamma_{121} \equiv \alpha/2, \quad \gamma_{112} \equiv \alpha^T/2, \quad \gamma_{222} \equiv&
\beta, \quad \gamma_{343} \equiv \alpha/2, \quad \gamma_{334} \equiv \alpha^T/2, \quad
\gamma_{444} \equiv \beta \;,
\end{split}
\end{equation}
and
\begin{equation}\label{eq:Omega}
 \Omega(k,\eta) \equiv \left(\begin{array}{cccc}
  0 & -1 & 0 & 0 \\
  -\frac32\Omega_m(1-f_\nu) & 1+{\cal H}'/{\cal H} & -\frac32\Omega_m f_\nu & 0
\\
  0 & 0 & 0 & -1 \\
  -\frac32\Omega_m(1-f_\nu) & 0 &  -\frac32\Omega_m(f_\nu-\frac{k^2}{k_{FS}^2})
& 1+{\cal H}'/{\cal H} 
 \end{array}\right)~.
\end{equation}
The dependence on $k$ enters via the free-streaming term $\frac{k^2}{k_{FS}^2}$, and
we used the notation $\alpha^T(k,p,q)\equiv \alpha(k,q,p)$. 

Within $\Lambda$CDM it is
convenient to further rewrite the evolution equations using the linear growth factor
as time variable. This allows one to approximately map them onto those for an EdS model with
good accuracy. In this case the time-dependence factorizes and the non-linear solution
can be expressed in terms of the well-known SPT kernels~\cite{Bernardeau:2001qr}.
This is not possible in presence of massive neutrinos due to the scale
dependence of the growth factor. Still, one can find a similar reparameterization
which we briefly describe in App.\,\ref{app:AltParam}. Nevertheless, these
equations cannot be mapped onto an EdS model to good accuracy, and therefore
we use a scheme that can account for the time- and wavenumber-dependent
entries $\Omega(k,\eta)$ described in the next section. Since there is no
advantage in performing a reparameterization we used the form
(\ref{eq:Omega}) in terms of conformal time $\eta=\ln(a)$ (or equivalently redshift) for the numerical solution. 

\subsection{Solution scheme \label{sec:sol}}

Even for the pure dark matter (DM) universe, the corrections to linear order in the standard perturbation theory (SPT) deviate from the $N$-body result significantly.
Considering the current bounds on the neutrino masses, non-linear effects from dark matter clustering alone are comparable in size to those from massive neutrinos
close to the BAO scale. Unfortunately, to obtain better precision in perturbation theory it is not enough to compute higher corrections in SPT since it is known that the SPT expansion is not convergent \cite{Bernardeau:2012ux,Blas:2013aba}. There are
currently two opposite views to deal with this problem. 
If one follows the results of  \cite{Pueblas:2008uv,Blas:2013aba} it seems that the divergence of the series is related to the unphysical treatment of SPT
 of modes well inside the regime of validity of the single fluid approximation. Thus, one expects that progress should come from a better understanding of the perturbative expansion. This is the idea behind  the resummation techniques of \cite{Crocce:2005xy,Taruya:2012ut,Matarrese:2007wc,Anselmi:2012cn,Matsubara:2007wj,Pietroni:2008jx,Blas:2013aba}.  
 An alternative possibility is to assume that the lack of convergence may disappear
 once the corrections to the perfect single-fluid approximation coming from short(non-linear)-modes are considered. 
To follow this idea, one can use an  effective description and extract information on the latter from
other sources (for example $N$-body simulations or data). This is behind the effective field theory of large-scale structure
 \cite{Carrasco:2013mua,Carroll:2013oxa,Mercolli:2013bsa,Porto:2013qua}.
Other approaches using effective descriptions beyond the perfect fluid are \cite{Pueblas:2008uv,Manzotti:2014loa,Pietroni:2011iz}.
Independently of which approach encapsulates the relevant physics, it is essential to be able to incorporate massive neutrinos in it\footnote{Different works have been  devoted to extend the non-linear techniques to other multi-fluid situations, in particular to account  for baryon or dark energy perturbations \cite{Somogyi:2009mh,Saracco:2009df,Brouzakis:2010md,D'Amico:2011pf,Anselmi:2011ef,Bernardeau:2012aq,Anselmi:2014nya}.}. 

Our purpose is to take a first step beyond linear order and solve the system (\ref{eq:PTinvectors}) at the one-loop level, without introducing approximations to the matrix $\Omega(k,\eta)$ of Eq.~(\ref{eq:Omega}).
This can be done following an approach introduced in Ref.~\cite{Pietroni:2008jx}, namely,
the one-loop limit of the time-flow equations. This approach was called {\em dynamical one-loop} in~\cite{Audren:2011ne}. The starting point is the set of 
flow equations obtained by
multiplying (\ref{eq:PTinvectors}) with several
equal-time fluctuation fields $\psi_a$, and then taking a statistical average. This
leads to a hierarchy of equations for the 
$n-$point correlation functions. By neglecting the 
connected four-point function (usually called $Q$), one obtains a closed system of evolution equations for the power spectrum $P$ and bispectrum $B$,
\bea
\label{eq:1loopdynP}
\partial_\eta P_{ab} (k,\eta) &=& 
 - \Omega_{ac}(k,\eta) P_{cb} (k,\eta)  - \Omega_{bc}(-k,\eta) P_{ac} (k,\eta) 
\nn \\
 && + \int \, \frac{d^3q}{(2\pi)^3} \left[ \gamma_{acd}(k,q,k-q) \, B_{bcd} (k,-q,q-k,\eta) \right. 
\nn  \\
 && \quad \left. +  \,  \gamma_{bcd}(k,q,k-q) \, B_{acd} (k,-q,q-k,\eta) \right] \,
,
\eea
and\footnote{Note the different convention $\gamma(k,p,q)\to \gamma(-k,p,q)$ when compared to~\cite{Pietroni:2008jx}.}
\bea
\label{eq:1loopdynB}
\partial_\eta B_{abc} (k,-q,q-k,\eta) &=& 
 - \Omega_{ad}(k,\eta) B_{dbc} (k,-q,q-k,\eta)  \nn \\
 && - \Omega_{bd}(-q,\eta) B_{adc} (k,-q,q-k,\eta)  \nn \\
 && - \Omega_{cd}(q-k,\eta) B_{abd} (k,-q,q-k,\eta)  \nn \\
 && + 2 \left[ \gamma_{ade}(-k,-q,q-k) \, P_{db} (q,\eta) \, P_{ec} (k-q,\eta) \right.  \nn 
\\
 &&    \quad \left. + \, \gamma_{bde}(q,q-k,k) \, P_{dc} (k-q,\eta) \, P_{ea} (k,\eta)
\right. \nn \\
 &&    \quad \left. + \,  \gamma_{cde}(k-q,k,-q) \, P_{da} (k,\eta) \, P_{eb} (q,\eta)
\right]  \, .
\eea
In the {\it dynamical one-loop} approach, the linear power spectrum is obtained by solving (\ref{eq:1loopdynP}) without the bispectrum as a source (or alternatively, by taking it directly from the output of a Boltzmann code). Next, the linear solution for the power spectrum $P$ is used in Eq.~(\ref{eq:1loopdynB}) to generate the leading solution for the bispectrum $B$. This bispectrum is finally used back in (\ref{eq:1loopdynP}) to yield the one-loop correction to the power spectrum $P$. For $\Lambda$CDM cosmologies, the final results do not deviate sizeably from the full time-flow solution, as well as the SPT one-loop calculation \cite{Audren:2011ne}. Still, compared to usual perturbation theory, this procedure has two advantages. First, these equations can be solved with reasonable numerical effort for arbitrary cosmologies and including a scale-dependence in the propagation, i.e. a $k$-dependent $\Omega$ in Eq.\,(\ref{eq:Omega}), or equivalently a $k$-dependent linear growth factor. Second, this scheme captures the well-known cancellations among the various perturbative contributions that occur in the limits $q\ll k$ and $|k-q|\ll k$ (soft loop momentum) as well as $k\ll q$ (soft external scale) \cite{Blas:2013bpa,Carrasco:2013sva}. Thus the terms on the right-hand side of Eq.\,(\ref{eq:1loopdynP}) are arranged in a way that is very suitable for numerical implementation. This is discussed in detail in Secs.~\ref{sec:IRcancellation} and \ref{sec:UVcancellation}, respectively. Furthermore, the method can be extended to higher orders by including further $n$-point correlations in the system.

If the scale dependence in $\Omega$ is neglected
only in the equation for the bispectrum, the system further
simplifies. If one is interested in the power spectrum, this approximation can be justified
because the bispectrum contributes only to the one-loop correction. More specifically, since the
linear neutrino power spectrum is already strongly 
suppressed below the free-streaming scale, any additional suppression in the neutrino propagation at the one-loop
level has only a marginal impact on the dark matter fluctuations. We discuss the impact of this approximation in detail
in Sec.~\ref{sec:bispectrum}. 

In this approximation, the propagation matrix $\Omega$ can be evaluated at the
external wavenumber $k$ in Eq.\,(\ref{eq:1loopdynB}), and 
the equations can be brought into the form~\cite{Pietroni:2008jx}
\bea\label{eq:flow}
\partial_\eta P_{ab} (k,\eta) &=& 
 - \Omega_{ac}(k,\eta) P_{cb} (k,\eta)  - \Omega_{bc}(k,\eta) P_{ac} (k,\eta) 
\nn \\
 && \quad + \, \left[ I_{acd,bcd}(k) +   I_{bcd,acd}(k)  \right] \, ,\nn\\
\partial_\eta I_{abc,def} (k,\eta) &=& 
 - \Omega_{dg}(k,\eta) I_{abc,gef} (k,\eta) - \Omega_{eg}(k,\eta) I_{abc,dgf}
(k,\eta)  \nn \\ 
&& - \Omega_{fg}(k,\eta) I_{abc,deg} (k,\eta) + 2 A_{abc,def}(k) \,,
\eea
where we follow the definitions in~\cite{Pietroni:2008jx},
\be
I_{abc,def}(k,\eta) \equiv \frac12 \int \, \frac{d^3q}{(2\pi)^3} \, \big( \gamma_{abc}(k,q,k-q) \, B_{def} (k,-q,q-k,\eta) \ + \   (q \leftrightarrow k-q) \big) \,,
\ee
and
\bea\label{eq:A}
A_{abc,def}(k,\eta) &\equiv& \frac12 \int \frac{d^3q}{(2\pi)^3} \, \frac{d^3p}{(2\pi)^3} \, (2\pi)^3 \delta(k-q-p) \left\{
\gamma_{abc}(k,q,p)  \right. \nn \\
&&  \times \left[ \gamma_{dgh}(k,q,p) \, P_{eg} (q,\eta) \, P_{fh} (p,\eta) \right.  \nn 
\\
 &&    \quad \left. + \, \gamma_{egh}(q,-p,k) \, P_{fg} (p,\eta) \, P_{dh} (k,\eta) \right.
\nn \\
 &&    \quad \left. + \,  \gamma_{fgh}(p,k,-q) \, P_{dg} (k,\eta) \, P_{eh} (q,\eta)
\right]  \nn \\
&& \quad \quad \left. + \quad  (p \leftrightarrow q) \quad \right\} \,.
\eea
This trick allows one to integrate the flow
equations using only one-dimensional functions which simplifies the numerical
problem drastically.

As mentioned above, the initial conditions are obtained at some finite redshift by matching the power spectra to the linear solutions ($P_{i,ab}=P_{ab}^{Boltzmann}$).
In perturbation theory higher orders are parametrically suppressed by an additional growing mode factor $D(z)^2 \sim a^2$. Therefore one naively expects that the relative error made by starting from the linear solution, and with a vanishing bispectrum, is of order $a_i^2 \sim z_i^{-2}$.
However, integrating the flow equations with an initially vanishing bispectrum
leads to an error of order $z_i^{-1}$. This 
is because the source $A_{abc,def}(k)$ in the equation of the bispectrum scales
as $a^4$ (for pure CDM and EdS cosmology) 
while the bispectrum itself, if not sourced, would evolve as $a^3$. 
For accurate predictions, it is therefore important to reduce the impact of the
initial conditions. In Ref.~\cite{Audren:2011ne} the bispectrum was initialized
with the leading perturbative contribution computed for an EdS cosmology, $B_{i}=B_{SPT}(z_i)$.
The impact of using such an initial condition instead of $B_{i}=0$ was found to be important
(leading to the conclusion that dynamical one-loop results and full flow equations provide very similar results). 

In our analysis, it would be non-trivial to compute an initial bispectrum analytically, due to neutrino effects.
We circumvent this issue by starting the time evolution for the bispectrum (second equation in \eqref{eq:flow})
at an earlier time $z_{ini} \sim z_i^2 = 25^2$. Within the interval $z_{ini}>z>z_i$
we use an extrapolated source term $A_{abc,def}^{extr}(k,z)\equiv (a/a_i)^4 A_{abc,def}(k,z_i)$ in terms of the power-spectrum at $z=z_i$. Later, at $z<z_i$, the proper source term $A$ is restored.  
The evolution between $z_{ini}$ and $z_i$ may be considered as a convenient way
to generate an `initial' condition for the bispectrum at $z_i$. For an EdS cosmology, the bispectrum at $z_i$ obtained in this
way coincides with the leading SPT result up to errors of order $z_i^{-1}$, such that the propagated error on the power spectrum at $z=0$ is reduced
to order $z_i^{-2} \lesssim 0.2\%$\footnote{We explicitly checked that our numerical implementation yields a power spectrum that agrees to this level with the SPT one-loop result for an EdS cosmology with CDM only.}. To be precise, the scaling assumed for $A_{abc,def}^{extr}(k,z)$ in the range $z_{ini}>z>z_i$ is not exact in presence of massive neutrinos, so the resulting total error is actually of order max$(f_\nu, z_i^{-1}) / z_i$. It is anyway at or below the permille level.

\section{Decoupling of scales in the exact solution}\label{sec:3}

\subsection{Cancellation for small wavenumber~\label{sec:IRcancellation}}

It is well known that in SPT, all loop effects in the power spectrum are in the
limit of small external wavenumber
$k$ suppressed by a factor $k^2$ compared to the linear spectrum.
However, this results from an intricate cancellation between different diagrams.

This cancellation can be traced back to the asymptotic behavior of the
interaction vertex $\gamma$. 
In a first step, let us neglect the neutrino fluid. In general one has
\be
\label{eq:ksoft}
\gamma_{abc}(k,q,p) \xrightarrow{k\to 0} \propto k \;.
\ee
Furthermore (neglecting the neutrinos) one has 
\be
\label{eq:gammaSoft}
\gamma_{abc}(k,p,q) \xrightarrow{q\to 0} \delta_{ab} \delta_{c2} \frac{k \cdot
q}{2 \, q^2} \, ,
\ee
and the symmetric result for $p\to 0$.
\begin{figure}
\begin{center}
\includegraphics[width=0.4\textwidth]{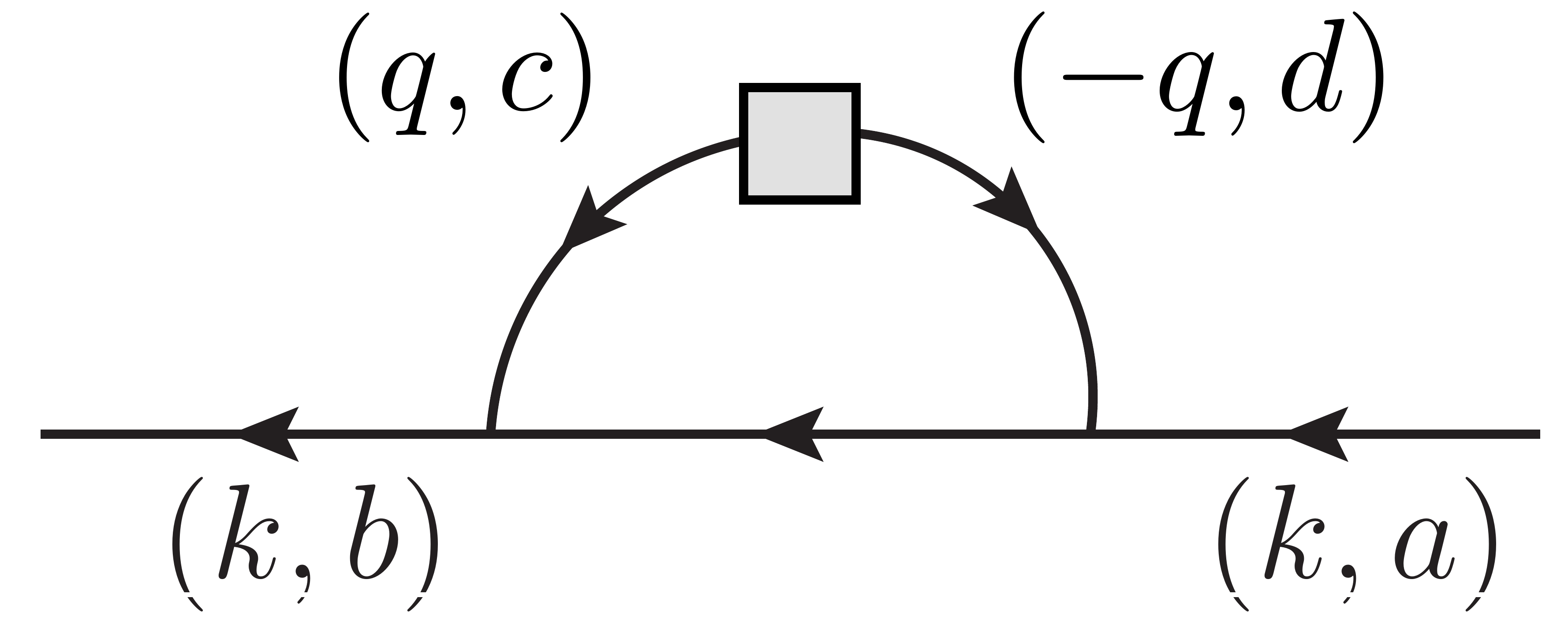}
\caption{\label{fig:SPTdiagrams}
The soft one-loop contribution to the propagator.}
\end{center}
\end{figure}

Now consider the one-loop correction to the propagator in Fig.~\ref{fig:SPTdiagrams}. From \eqref{eq:ksoft}, the 
vertex of the external leg $(k,b)$ scales as $k$. At the other side of the diagram, the incoming line $(k,a)$  ends in a
vertex that from  \eqref{eq:gammaSoft} 
leads to a potential enhancement of order $1/k$.
So, naively the one-loop correction to the propagator is not suppressed compared to the linear
one for $k \to 0$. 
However, this is not true because the effect of adding the incoming soft line can
be factorized: the whole structure associated to the vertex of the line $(k,a)$ can be expressed as a term multiplying
the propagator from the initial power spectrum to the vertex connecting to $(k,b)$
 (here $g_{ab}(\eta_1, \eta_2)$ denotes the linear propagator)
\bea\label{eq:factorize}
g_{fc}(\eta_1, \eta_c) \gamma_{cda}(k-q,-q,k)
g_{de}(\eta_c, \eta_2)  
&\simeq&
 g_{fc}(\eta_1, \eta_c) g_{ce}(\eta_c, \eta_2) \delta_{a2} \frac{k \cdot
(k-q)}{ 2\, k^2} \nonumber\\  
&\simeq& - g_{fe}(\eta_1, \eta_2) \delta_{a2} \frac{k \cdot q}{2 \,k^2}\, .
\eea
This relation is essential for the factorization of soft effects.

Since the momentum $q$ should be integrated for positive and negative values the leading term in fact
cancels. This can also be understood as the contribution coming from attaching the soft line to the $(q,c)$ leg in Fig. \ref{fig:SPTdiagrams}:
\bea
\textrm{ one-loop propagator} \quad &\propto& \quad
\frac{k\cdot q}{2q^2} \quad \times 
\left\{ \frac{k \cdot q}{2\, k^2} + \frac{k \cdot (-q)}{2\,k^2} +{\cal O}(k^0)\right\} \nn \\
&\propto& \quad \textrm{linear propagator} \times k^2
\eea
The last relation follows from the fact that because of isotropy, the propagator
cannot be linear in the external wavenumber $k$. Note that, if the $1/k$ contributions did not cancel,
then a contribution proportional to the linear propagator\ $\times\ k^0$ would be generated. 

This argument applies not just to the one-loop diagram but to general diagrams contributing to the propagator.
Attaching the incoming line of wavenumber $k$ to a propagator carrying wavenumber $q_i$
of a given diagram gives an enhancement of order $ k \cdot q_i /
k^2$. When summing over all possibilities, the factorization ensures that the 
contributions that scale as $1/k$ cancel in the sum~\cite{Blas:2013bpa}.

Now consider the same situation including the neutrinos. The vertex behaves as 
\be
\label{eq:softmulti}
\gamma_{abc}(k,p,q) \xrightarrow{q\to 0} \delta_{ab} 
\left[ (\delta_{a1} + \delta_{a2})\delta_{c2}  +  
(\delta_{a3} + \delta_{a4})\delta_{c4} \right] 
\frac{k \cdot q}{ 2  \, q^2} \, ,
\ee
where the indices correspond to those in \eqref{eq:PTinvectors}.
This does not {\em per se} become diagonal in the soft limit. However, soft
fluctuations are dominated by the linear growing mode such that
\be
\label{eq:softmulti2}
\gamma_{abc}(k,p,q)\psi_c(q) \xrightarrow{q\to 0} -\delta_{ab} 
\frac{\theta(q)}{{\cal H}}\frac{k \cdot q}{ 2  \, q^2} \, ,
\ee
due to $\theta(q) \simeq \theta_\nu(q)\simeq \theta_{cb}(q)$ for $q\to 0$.
Therefore, soft effects factorize in perturbative
calculations and the same cancellation of soft 
enhancement effects occurs when including neutrinos. 

Let us compare this to schemes that treat the neutrinos linearly, acting as a background for the CDM perturbations (see
e.g.~\cite{Wong:2008ws,Lesgourgues:2009am}). This corresponds in
our framework to setting the neutrino sector of the interaction to zero. Once
this approximation is done, the soft effects are not proportional to $\delta_{ab}$  as in (\ref{eq:softmulti2}), such that
the factorization of the soft incoming line in (\ref{eq:factorize}) does not occur any more.
The cancellation of soft effects is 
 incomplete and the loop-contributions to the power spectrum are not 
suppressed for small external wavenumbers.  This also has a sizeable 
numerical impact on the one-loop contribution in the BAO regime (see discussion
in sec.~\ref{sec:app_vs_full}). Furthermore, this approximation completely 
breaks down on the two-loop level, since the two-loop integrals become UV
divergent when the relative $k^2/q^2$ suppression of short modes in the momentum integrals is absent. This is even true for large red-shifts where 
cosmological perturbation theory is converging quickly.

\subsection{Cancellation for soft loop momenta~\label{sec:UVcancellation}}

As has been demonstrated e.g. in \cite{Pietroni:2008jx}, the cancellations which occur in the opposite limit, namely for
soft loop momentum $q\ll k$ (compared to the external wavenumber $k$), are implemented within
the flow-equation framework in a way which is well suited for numerical implementation.
Here we briefly point out that this finding persists when including massive neutrinos.
Before that, let us briefly summarize the situation in SPT and within the dynamical one-loop
scheme for CDM only. Specifically, the scaling $\gamma \propto k/q$ of the
vertices potentially leads to large contributions to the non-linear corrections.
Within the usual SPT framework, this yields large contributions to individual terms (e.g. $P_{22}$ and $P_{13}$ in the usual notation)
which are of opposite sign and cancel in the sum. Especially when going to higher loop orders, this is challenging for a numerical evaluation
and additional care is needed~\cite{Scoccimarro:1995if,Blas:2013bpa, Blas:2013aba,Carrasco:2013sva} to ensure all cancellations. 

Notably, within the dynamical one-loop described here, the corresponding cancellations occur already at the level of the integrand from the outset. For $p \to 0$ the first and second interaction terms in the square bracket in Eq.\,(\ref{eq:A}) cancel in leading order in $k/p$, and in the limit $q \to 0$ the first and the third term cancel in leading order in $k/q$. This is due to the fact that the flow equations only contain equal time correlators, unlike usual perturbation theory. This cancellation of soft effects also ensures the correct behavior for the bispectrum in the squeezed limit~\cite{Peloso:2013zw,Kehagias:2013yd, Creminelli:2013mca,Valageas:2013cma, Kehagias:2013paa,Horn:2014rta}. The cancellation on the integrand level extends to multi-fluid systems as long as one considers dominant growing mode perturbations (or more generally adiabatic modes with $\theta_{\nu}/\theta_{cb}\to 1$) for the modes $k\to 0$ at each order~\cite{Bernardeau:2012aq}. In the present context, we find that the cancellations occur when treating neutrinos non-linearly. Additionally, in contrast to the opposite regime $k\ll q$ discussed previously, the cancellations for soft loop momenta even occur when treating neutrinos linearly as a background source for the CDM perturbations~\cite{Wong:2008ws,Lesgourgues:2009am}.

\subsection{Decoupling of scales and momentum conservation}\label{sec:dipole}

The cancellations in the large-scale limit discussed in Sec.\,\ref{sec:IRcancellation} lead to
a screening of the impact of small-scale perturbations on non-linear corrections \cite{Goroff:1986ep}. This is not accidental but
can be related to the conservation of total momentum 
of the system and its consequences for the gravitational interaction \cite{Rees:1982,Peebles:1980}\footnote{A related but distinct symmetry argument is also
behind the recently derived consistency relations for large scale structure \cite{Peloso:2013zw,Kehagias:2013yd,Creminelli:2013mca,Valageas:2013cma,Kehagias:2013paa,Horn:2014rta}, which result from a screening of the impact of \emph{large-scale} perturbations.}. Therefore, it is important to
employ an approximation scheme that is compatible with this conservation  law in order to ensure
the cancellations in the large-scale limit. When the neutrino component is treated linearly, it enters the fluid equations for the CDM/baryons as
an external source instead of being computed self-consistently. This potentially spoils
total momentum conservation, because momentum can be transferred into the neutrino sector. To
restore total momentum conservation it is therefore necessary to take the corresponding
back-reaction on the neutrinos into account, which is ensured for example when treating
both fluid components on an equal footing. We devote the rest of this section to elaborate
on the relation between momentum conservation and decoupling of scales 
(and further screening appearing in virialized structures). 
 
We first recall the connection between the cancellations discussed in Sec.\,\ref{sec:IRcancellation}
with the overall momentum conservation. For the sake of the argument, it is helpful to consider
an artificial initial condition that features fluctuations only on small scales, say for $k>\Lambda$
with some (arbitrary) cutoff scale $\Lambda$. Then one may consider the perturbations that are
generated with time on large scales $k\ll \Lambda$ via the nonlinear coupling to the short modes.

\begin{figure}
\begin{center}
\includegraphics[width=0.65\textwidth]{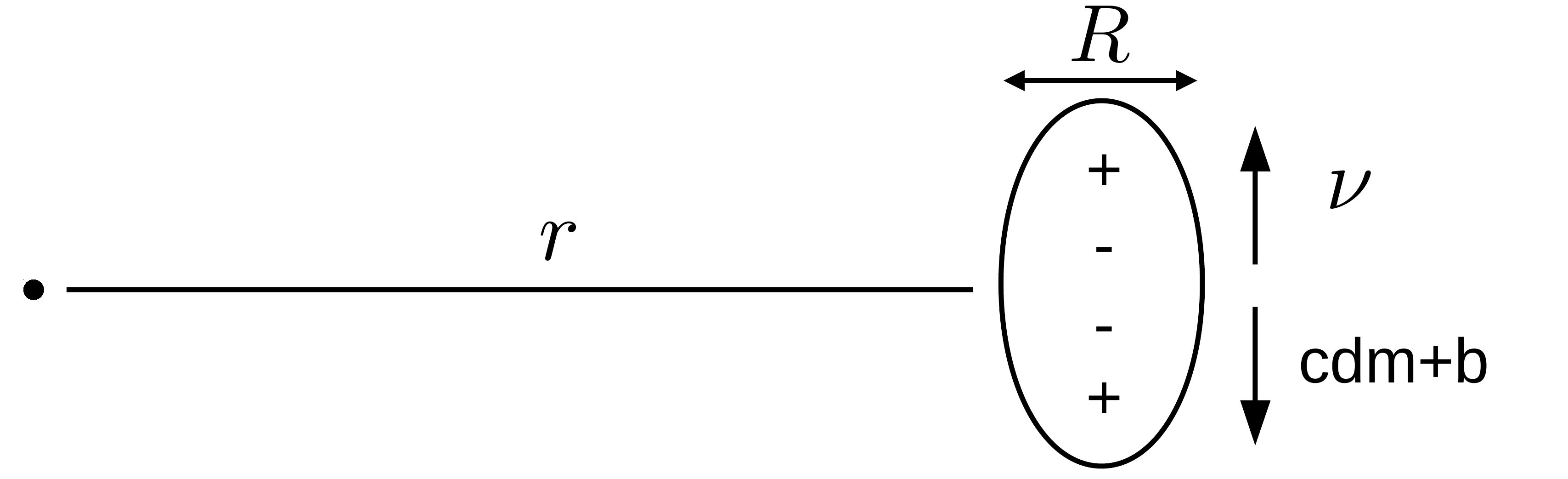}
\end{center}
\caption{\label{fig:quadrupole} Illustration of the screening of short-scale fluctuations
on large scales. When treating neutrino and CDM/baryon perturbations on an equal footing,
momentum conservation ensures that the leading contribution at large distances is the
quadrupole moment. Within approximation schemes that treat neutrinos and CDM/baryons
differently (e.g. neglecting non-linearities for the former) potentially a
spurious dipole contribution can be generated which upsets the $k^2$-suppression of
non-linear corrections to the matter power spectrum at small $k$. 
}
\end{figure}

One may gain some intuition by considering an even simpler setup where the small-scale
fluctuations are also spatially confined to a region $R \sim 1/\Lambda$. Then the impact on large distances
can be understood in terms of the multipole expansion of the gravitational potential \cite{Rees:1982}. Indeed,
since $\delta$ has zero mean (a property that is preserved under time evolution due to mass conservation), the
monopole vanishes. In addition, local momentum conservation on scales $\sim R$ forbids also the formation of
a dipole, leaving the quadrupole moment as the leading contribution, which leads to a suppression of the gravitational
potential $\phi(r) \propto R^2/r^3$ for $r\gg R$ (see Fig.\,\ref{fig:quadrupole} for an illustration).

The same effect can be seen by considering the Fourier transform of the total density field $\delta\equiv \sum_A \delta_A$ (see \eqref{eq:totaldelta}), 
\be
\delta_k(\tau)= \int d^3 x \,\delta(x,\tau) e^{-i k \cdot x}\,.
\ee
To obtain the large-scale limit $k\ll \Lambda$ we can expand the exponential
\be\label{eq:DeltaTaylor}
\delta_k(\tau)= \int_R d^3 x\, \delta(x,\tau) - ik^j  \, \int_R d^3 x\, x^j\delta(x,\tau) +{\cal O}(k^2)\,.
\ee
The first term cancels by definition initially, and also at later times due to mass conservation, analogous
to the monopole discussed above. The second term can be related to the total momentum. 
Using the continuity equation  to evaluate the first derivative w.r.t the conformal time $\tau$
one obtains the total momentum,
\be
\label{eq:CM}
\frac{d}{d\tau}\int_R d^3 x\, x^i\delta(x,\tau)=\sum_A\int_R d^3 x\rho(x,\tau)_A\, v^i(x,\tau)_A/\bar \rho(\tau)\,,
\ee
where $A$ labels the fluid components (i.e. $A=cb,\nu$).
The time evolution of the total momentum can be obtained using both continuity and Euler equations, 
and a straightforward computation yields (we assume that the background evolution is dominated by matter, but 
this is not essential) 
\be
\begin{split}
\frac{d}{d\tau} \sum_A\int_R d^3 x\rho_A\, v^i_A=&-4\HH  \sum_A\int_R d^3 x\rho_A\, v_A^i\\
&-\int_R d^3 x\sum_A\Big( v_A^i\partial_j(\rho_A v_A^j)
+\rho_A v_A^j\partial_j v_A^i+\rho_A \partial_i\phi\Big),
\end{split}
\ee
where we omitted the pressure terms since they would anyway yield corrections of ${\cal O}(k^2)$ in  \eqref{eq:DeltaTaylor}.
The first term on the right-hand side just describes the usual dilution due to redshift.
The second and third terms can be combined into a total derivative, such that they vanish after spatial
integration\footnote{For the artificial setup we consider at the moment there is no flux outside the region of
size $R$ because we assume that perturbations vanish outside, i.e. it is isolated. We comment on the implications for
a more realistic case below.}. Note that they originate from
different contributions of the continuity and Euler equation, and that it is important to keep
both linear and non-linear pieces.
Using the Poisson equation, the final term can also be rewritten as a total derivative
\be
\label{eq:phi}
 \int d^3 x\sum_A\rho_A \partial_i\phi = -\frac{1}{4\pi G}\int d^3x \, \Delta \phi \, \partial_i \phi
 = \frac{1}{8\pi G}\int d^3x \, \partial_i (\nabla \phi )^2 = 0 \;.
\ee
Thus, as expected the total momentum is conserved up to the pressure contribution and standard dilution with the scale factor,
\be
\label{eq:gravp}
\sum_A\int d^3 x\rho_A\, v_A^i=\sum_A\int d^3 x\rho(t_0)_A\, v_A^i(t_0)/a^4.
\ee
Since we can initially move to the rest frame of the fluid, this implies the familiar result that the center of mass
stays at a fixed position (cf. \eqref{eq:CM}),
\be
\int_R d^3 x\, x^i\delta(x)=C^i\,.
\ee
Using the freedom to choose a coordinate frame where the center of mass is at the origin\footnote{The same result can be obtained in any frame after statistical averaging.}, we finally find that the long-scale perturbations induced by the short fluctuations are suppressed by $k^2$,
\be
\delta_k \sim k^2.
\ee
This leads to the well-known result that the long-range `tail' of the power spectrum  generated purely by the fluctuations of short modes
scales as\footnote{Note that the same result also holds in the kinematic description of the system.} $P \propto k^4$ \cite{Zeldovich:1965,Peebles:1980,Fry:1993bj,Padmanabhan:2005vb}. When applied to a single-fluid in an EdS background, this can be directly related to the `decoupling' property of
the SPT kernels, namely that 
\be\label{eq:decoupling}
  F_n(k_1,\dots, k_n) \propto k^2/q^2,
\ee
where $k=\sum_ik_i$ and it is assumed $k_i \sim q$, $|k|\ll |q|$. For example, at the
one-loop level, for the particular setup considered here one has $P_{13}(k)=0$ for $k<\Lambda$ because it is proportional to the initial power
spectrum evaluated at $k$, which is assumed to vanish. Therefore, the leading contribution is given by
\be
  P_{22}(k) = \int d^3q F_2(q,k-q)^2 P_{lin}(q) P_{lin}(k-q) \propto k^4 \int d^3q P_{lin}(q)^2/q^4 \,,
\ee
where we used the `decoupling' property for $F_2$. In fact, it is easy to see that the same argument holds for all loop corrections $P_{mn}$: if $m=1$ or
$n=1$ one has $P_{mn}=0$ for $k<\Lambda$, analogously to $P_{13}$, while otherwise $P_{mn}\propto k^4$ by virtue of the `decoupling' property for $F_n$ and
$F_m$, respectively. This agrees with the expectation from the argument above based on total momentum conservation, and therefore we expect that the `decoupling' property holds also in more general settings provided the underlying approximation is compatible with the overall conservation of momentum. At this point it is worth noting that the scaling $F_n\propto k^2$ is due to intricate cancellations among various contributions in the perturbative calculation, e.g. based on the well-known recursion relations, especially for large $n$ \cite{Goroff:1986ep}. Based on the arguments above we expect that these cancellations are spoilt within approximation schemes that are not compatible with total momentum conservation.

Let us now comment on the implications for a realistic setup where perturbations are present initially on all scales, with $P_{lin} \propto k^{n_s}$, $n_s\simeq 1$.
Since the kernels are independent of the initial conditions, the property (\ref{eq:decoupling}) is generally valid. However, in contrast to the previous case, $P_{13}$ dominates over $P_{22}$ for small $k$ for the realistic setup. Nevertheless, the `decoupling' property for $F_3$ then yields the well-known scaling $P_{13} \propto k^2 P_{lin}(k)$. Analogously, the same argument applies to all $P_{1n}$.

The previous argument is violated when neglecting non-linearities for the neutrino component, since in that case Eq.~\eqref{eq:phi} does not reduce to a total derivative. 
More concretely, if one considers the non-linear evolution equations for the CDM/baryon components only, and takes into account the (linear) neutrino perturbations solely via their contribution to the gravitational potential in the Poisson equation, Eq.~\eqref{eq:CM} has an extra contribution such that (up to the dilution term)
\be
\frac{d^2}{d\tau^2}\int_R d^3 x\, x^i\delta_{cb}(x)=\frac{3\Omega_m(\tau) \HH^2}{8\pi}f_\nu\int d^3 x d^3\tilde x\,\delta_{cb}(x)  \delta_\nu^{lin}(\tilde x) \,\frac{(x^i-\tilde x^i)}{|x-\tilde x|^3}.
\ee
This contribution is generated because there is a net momentum lost in 
the description. Momentum conservation would be restored when considering the total density contrast $\delta=(1-f_\nu)\delta_{cb}+f_\nu\delta_{\nu}$, \emph{and} taking the non-linear
terms for the neutrino into account. If the latter are neglected, a leading ${\cal O}(k)$ behaviour is generated even for the total density contrast in Eq.\,(\ref{eq:DeltaTaylor}). 
This spurious term is very important at low-$k$ even for small $f_\nu$. In particular, this implies that the decoupling property (\ref{eq:decoupling}) is violated. For the case with realistic initial conditions this has the consequence that the relative suppression $P/P_{lin}\propto k^2$ does not occur, and that the sensitivity to UV modes inside the loop integrals is strongly enhanced. In fact, it turns out that this effect is very big for calculations beyond one-loop order, as has been mentioned in Sec.\,\ref{sec:IRcancellation}.
This can be understood by looking e.g. at $P_{15}=15P_{lin}(k)\int d^3p d^3q F_5(k,p,-p,q,-q) P_{lin}(p)P_{lin}(q)$, which is the dominant two-loop contribution on large scales. To estimate the sensitivity to UV modes, one may assume that the two power spectra inside the integral scale like $p^n$ for large $p$. Then, using that $F_5\propto k^2/\max(p^2,q^2)$ for $k\ll p,q$, the superficial degree of divergence when $p$ and $q$ become large is $D=2\cdot 3+2\cdot n-2$. For a realistic $\Lambda$CDM spectrum $n\sim -3$ and $D<0$, i.e. the integral is convergent. If the neutrinos are treated linearly, such that the decoupling property is violated, one will obtain a similar structure of the loop integral. However, as discussed above the kernel will not have a $k^2/\max(p^2,q^2)$ suppression, but instead scale like ${\cal O}(\max(p,q)^0)$. Therefore the superficial degree of divergence is in this case $D=2\cdot 3+2\cdot n$. Consequently, for the $\Lambda$CDM case one has $D\sim 0$, i.e. the loop integral is logarithmically UV divergent. Since the slope $n$ is even larger on moderately non-linear scales, this can have a large impact on the result for the two-loop result even if an ad hoc cutoff at the non-linear scale is introduced.

\subsubsection{Decoupling of virialized structures}

The decoupling of short-modes in $\Lambda$CDM is even more efficient for virialized structures \cite{Peebles:1980,Baumann:2010tm}. In fact, in this
case even the ${\cal O}(k^2)$ in Eq.\,(\ref{eq:DeltaTaylor}) can be seen to cancel due to momentum conservation. Given the previous discussion it is natural to wonder whether this 
complete screening is also absent if neutrinos are not consistently considered. 
To see how this happens\footnote{As before, an equivalent 
derivation can be done in the kinetic picture.}, we consider a  time interval $T$ small with respect to the ${\cal H}^{-1}$. In this approximation
\be
\label{eq:tensorvirial}
\frac{d}{d \tau}\sum_A\int d^3 x\rho_A v_A^jx^i=\sum_A \int d^3 x\rho_A v_A^jv_A^i-\int d^3 x\rho x^i
\partial_j \phi.
\ee
Assuming that the system is virialized (in the sense that the integrand in the l.h.s does not grow polynomially with $T$), the time-average of the l.h.s 
vanishes for large $T$, which yields the tensor virial theorem. Considering now the equation for the total $\delta$ (as before, we ignore all effects related to times $\HH^{-1}$) 
\be
\label{eq:NLk}
\bar \rho\,\partial_\tau^2 \delta_k=\int d^3 x\, e^{-ik x}\left[\partial_i(\rho \partial_i \phi)+\sum_A\partial_j \partial_i (\rho_A v_A^i v_A^j)\right]\,.
\ee
Let us divide $x=x_{CM}+x_p$, where $x_{CM}$ is the center of mass of the virialized structure (that we assume to be isolated in a  radius $R$, 
see Fig.~\ref{fig:quadrupole}). 
To understand  the effect of the non-linear terms in Eq.~\eqref{eq:NLk} at low-momentum, let us take the limit   $k R\ll 1$ and expand the r.h.s. of \eqref{eq:NLk},
\be
\label{eq:exp}
 \int d^3 x \big( i k_i(\rho \partial_i \phi) + k_i k_j \sum_A\rho_A v_A^i v_A^j\big)(1+i k (x-x_{CM})+{\cal O}(k^2)).
\ee
Using the Poisson equation, the ${\cal O}(k)$ contribution vanishes (cf. Eq.~\eqref{eq:phi}).
The ${\cal O}(k^2)$ term reads
\be
\label{eq:virialcan}
  k_i k_j \int d^3 x \big(\sum_A\rho_A v_A^i v_A^j-\rho (x-x_{CM})^j\partial_i \phi \big).
\ee
Using again the Poisson equation and the fact that $x_{CM}=$const., the last term in the last expression can be shown to be identical to the last term in Eq.~\eqref{eq:tensorvirial}, which
implies the cancellation of the ${\cal O}(k^2)$ corrections in \eqref{eq:NLk}. 

If non-linearities in the neutrino contribution are neglected, one finds for the CDM/baryon component
\bea
\int d^3 x \rho_{cb} (x-x_{CM})^j\partial_i \phi &=& \int d^3 x \rho_{cb} x^j\partial_i \phi  \nn \\
&& \hskip -3 cm -\frac{3\Omega_m \HH^2f_\nu}{8\pi}\int d^3 x d^3\tilde x\, \rho_{cb}(x)  \delta^{lin}_{\nu}(\tilde x) \frac{x^j(x^i-\tilde x^i)}{|x-\tilde  x|}.
\eea
To restore the cancellation of ${\cal O}(k^2)$-contributions one would have to sum over CDM/baryon and neutrino components, \emph{and} take non-linearities of the neutrinos into account, as before. Thus, we see again how treating neutrinos linearly can lead to spurious effects that are not present in the exact case.

\subsubsection{Dipole perturbations in alternative scenarios}

As a side remark, let us note that the previous discussion can be turned very useful for certain non-standard cosmological scenarios, for which the
conditions that ensure the absence of dipole perturbations are actually violated, such that the screening is incomplete (e.g. by violating the equivalence principle in DM or coupling the latter non-universally to a long-range fifth force). In those cases, the back-reaction of small scales on large scales is enhanced as compared to the standard picture, which suggests that these non-linear effects may be used to constrain certain deviations from $\Lambda$CDM\footnote{This
is similar to what happens in the emission of gravitational waves by binary systems where 
dipolar radiation is forbidden in general relativity but allowed in modified theories that violate any form of the equivalence principle.
Since dipolar radiation is very efficiently emitted (enhanced by a factor ${\cal O}(c^2/v^2)$ relative to the quadrupolar radiation, where $v$ is the characteristic orbital velocity),  this allows one to put very strong bounds on these theories \cite{Will:2014xja,Yagi:2013ava}.}.
Similar ideas have been studied for the consistency relations in \cite{Peloso:2013spa,Creminelli:2013nua}. 

Indeed, for any scenario with a modified Poisson equation of the form
\be
\Delta \phi=\frac{3}{2}{\cal H}^2\sum_i\Omega_i \delta_i+S(x),
\ee
with $S(x)$ parameterizing the deviations w.r.t standard gravity, or where the force term in the Euler equations is
 component-dependent or includes an extra force that is not sourced by the total density,
the Eq.~\eqref{eq:phi} and Eq.~\eqref{eq:virialcan} do not cancel. 
As a consequence, one has a double enhancement of the back-reaction effects (a first enhancement coming from the ${\cal O}(k)$ term  in Eq.\,(\ref{eq:DeltaTaylor})\footnote{In fact, the enhancement is even of ${\cal O}(k^0)$ if the power-spectrum already has
soft modes, as shown in Sec.~\ref{sec:IRcancellation}.} and another one at ${\cal O}(k^2)$ from virialized structures).
Thus, one may expect that these and similar effects related to the non-linear structure of the cosmological equations 
and which are suppressed in $\Lambda$CDM may be relevant to put bounds on modified gravity, 
the existence of extra-forces or certain violations of the equivalence principle. 

As a final comment we would like to clarify that 
even if the `momentum' defined in Eq.~\eqref{eq:gravp} is not conserved, the theory can still be translational invariant, with a conserved total momentum which includes contributions from an extra force, or which needs to be defined differently (see e.g. \cite{Yagi:2013ava,Blas:2012vn} and references therein for related discussion and concrete examples). As an illustration, we mention the possibility that the active and passive masses are different for the different  components \cite{Will:2014xja}. In this case, the Euler equation for the  component $A$ is modified as 
\be
\partial_\tau v_A ^i+{\cal H} v_A^i+v_A ^j \partial_j v_A ^i=-\alpha_A \partial_i \phi,
\ee
which means that, if the Poisson equation is not modified, the conserved momentum is
\be
\sum_A\int d^3 x \rho_A\, v_A^i/\alpha_A .
\ee
From now on we return to the discussion of $\Lambda$CDM with massive neutrinos which is the main focus of our work.

\section{Numerical results}\label{sec:4}

Before presenting our main results obtained from the two-fluid flow equations, and comparing them to various approximate one-fluid schemes, we briefly discuss the underlying linear treatment of the neutrino distribution. 

Throughout this work, in numerical applications, we use the parameter settings $M_\nu=\sum m_\nu = 0.21$\,eV, $h=0.72$, $\Omega_m^0=0.26$, $\Omega_b^0=0.044$, $n_s=0.96$, $A_s=2.385\cdot 10^{-9}$, $k_{pivot}=0.002\, h/$Mpc. For simplicity we assume three degenerate neutrino species. Note that we adjust the CDM density according to $\Omega_c^0=\Omega_m^0-\Omega_b^0-\Omega_\nu^0$, where $\Omega_\nu^0=f_\nu\Omega_m^0$. For the adopted set of parameters, one has $f_\nu\simeq 0.0167$ and $k_{FS}\simeq 0.032 (1+z)^{-1/2} h/$Mpc. All numerical results have been checked with two different C codes, written independently by two of us, and always found to be in good agreement.

\subsection{Linear solution}\label{sec:linear}

\begin{figure}
\begin{center}
\includegraphics[width=0.6\textwidth]{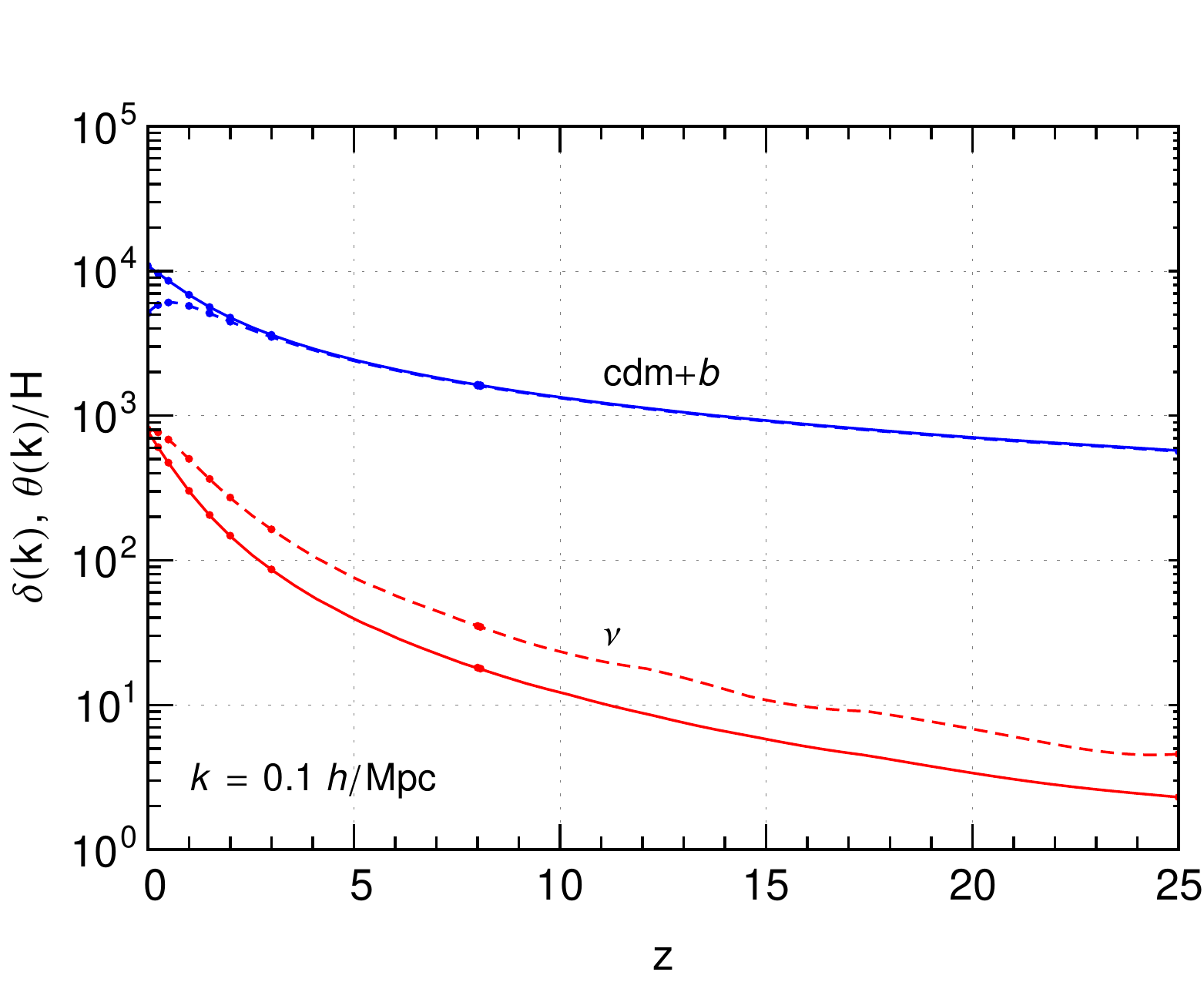}
\end{center}
\caption{\label{fig:lin1} Linear solution for three neutrino species with $M_\nu=\sum
m_\nu=0.21\,$eV as function of redshift for $k=0.1 h/$Mpc. The solid blue line corresponds to $\delta_{cb}$,
and the solid red line to $\delta_{\nu}$. The dashed lines show the velocity divergence $-\theta_{cb}/{\cal H}$ (red)
and $-\theta_\nu/{\cal H}$ (blue). The dots are taken from the solution of the Boltzmann equations obtained with CLASS.
}
\end{figure}

\begin{figure}
\begin{center}
\includegraphics[width=0.6\textwidth]{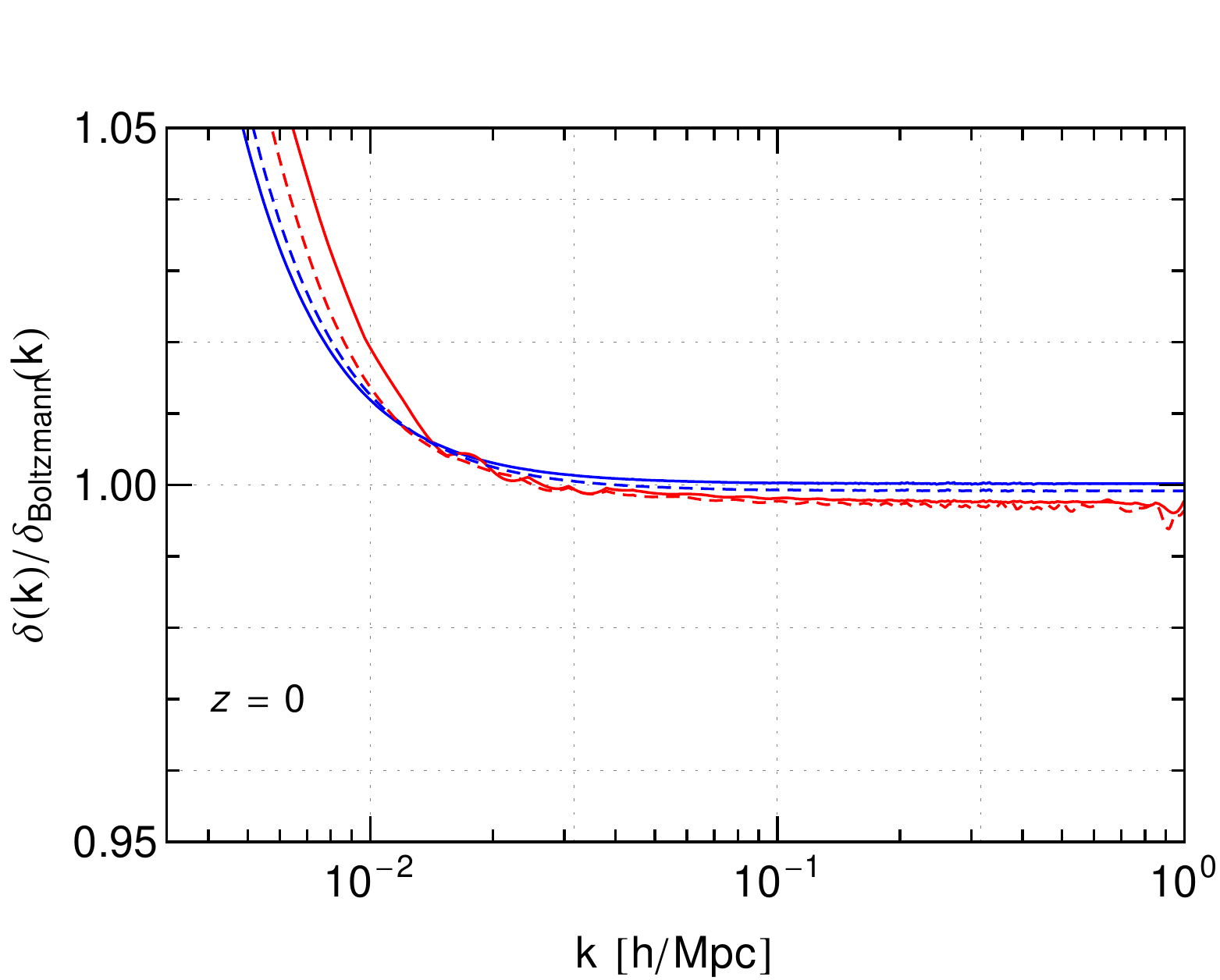}
\end{center}
\caption{\label{fig:lin2} Linear solution for three neutrino species with $M_\nu=\sum
m_\nu=0.21\,$eV as function of $k$ for $z=0$. The solutions are normalized to the corresponding solution of the Boltzmann equations obtained from CLASS. The assignment of the lines is as in Fig.\,\ref{fig:lin1}. The deviations for small $k$ are not related to the fluid approximation, but signal the influence of (gauge-dependent) relativistic corrections to the Newtonian limit.}
\end{figure}

As described in Sec.\,\ref{sec:2}, we use the full Boltzmann hierarchy for $z>25$, and a two-fluid scheme for $z<25$. As a first step, we checked the 
accuracy of this scheme at the linear level. We compared the density contrast and velocity divergence obtained from the linearized two-fluid equations at $z<25$ to the full Boltzmann solution obtained with CLASS \cite{Lesgourgues:2011re,Blas:2011rf} with high-precision settings, see Figs.~\ref{fig:lin1} and \ref{fig:lin2}. We find that the density contrast and the velocity divergence agrees to ${\cal O}(0.1\%)$ for the CDM, and to ${\cal O}(1\%)$ for the neutrino component, at $k\gtrsim 0.03 \,h/$Mpc. The differences at smaller $k$ are not related to the truncation of the Boltzmann equation, but due to deviations from the Newtonian limit on large scales close to the Hubble radius\footnote{\label{footnote:sync}This issue could be largely avoided by using the synchronous instead of Newtonian gauge for the full Boltzmann solution, because in this gauge, modes close to the horizon scale have a similar growth factor as subhorizon modes. However, since we want to match also the velocity divergence of the CDM component at $z=z_{match}$, we chose to adopt the Newtonian gauge.}. 
At first sight, these deviations from the Newtonian limit seem to hamper any predictions at large scales. Nevertheless, we stress that the linear solution is used here as an input to compute the one-loop integrals. Even for small external wavenumbers, the loop integrals are dominated by wavenumbers $q\gtrsim 0.05\, h/$Mpc for a realistic power spectrum, and therefore the deviations are not propagated to the non-linear corrections. To be more specific, to minimize the impact of deviations from the Newtonian limit, we compute the power spectrum using $P=P_{lin}^{Boltzmann}(1+\Delta P/P_{lin}^{fluid})$, where $\Delta P$ denotes the non-linear correction computed via the dynamical one-loop equations based on the linear spectrum obtained from the fluid equations $P_{lin}^{fluid}$. Since $\Delta P(k) \propto P_{lin}^{fluid}(k)$ for small $k$ the deviations cancel to a good accuracy, and the result is proportional to $P_{lin}^{Boltzmann}$ taken directly from the Boltzmann code. For large $k$, on the other hand, both linear spectra agree well such that the non-linear term is correctly reproduced as well. Additionally, the deviations in the neutrino component lead to an error of the order $f_\nu\times {\cal O}(1\%)$ in the non-linear results, which is negligibly small.

\subsection{Nonlinear solution}

To compute non-linear corrections, we solve the flow equations~(\ref{eq:flow}) for a two-fluid system composed of the neutrino component and
a pressureless CDM/baryon component. This scheme captures the time- and scale-dependent free streaming scale, and takes into account non-linearities in both components in a consistent manner. As explained above, this is important in order to obtain the correct scaling of the non-linear contributions in the $k\to 0$ limit, in accordance with momentum conservation and the expected cancellation of soft effects. In the following we first present our two-fluid solutions, and then discuss the
validity of the approximate treatment of the bispectrum on which our analysis is based (as well as all previous analyses known to us).
Finally, we compare to various approximation schemes, with special emphasis on the errors introduced by sacrificing the consistency of the two-fluid scheme.

\subsubsection{Numerical results for the two-fluid scheme}

\begin{figure}
\includegraphics[width=0.49\textwidth]{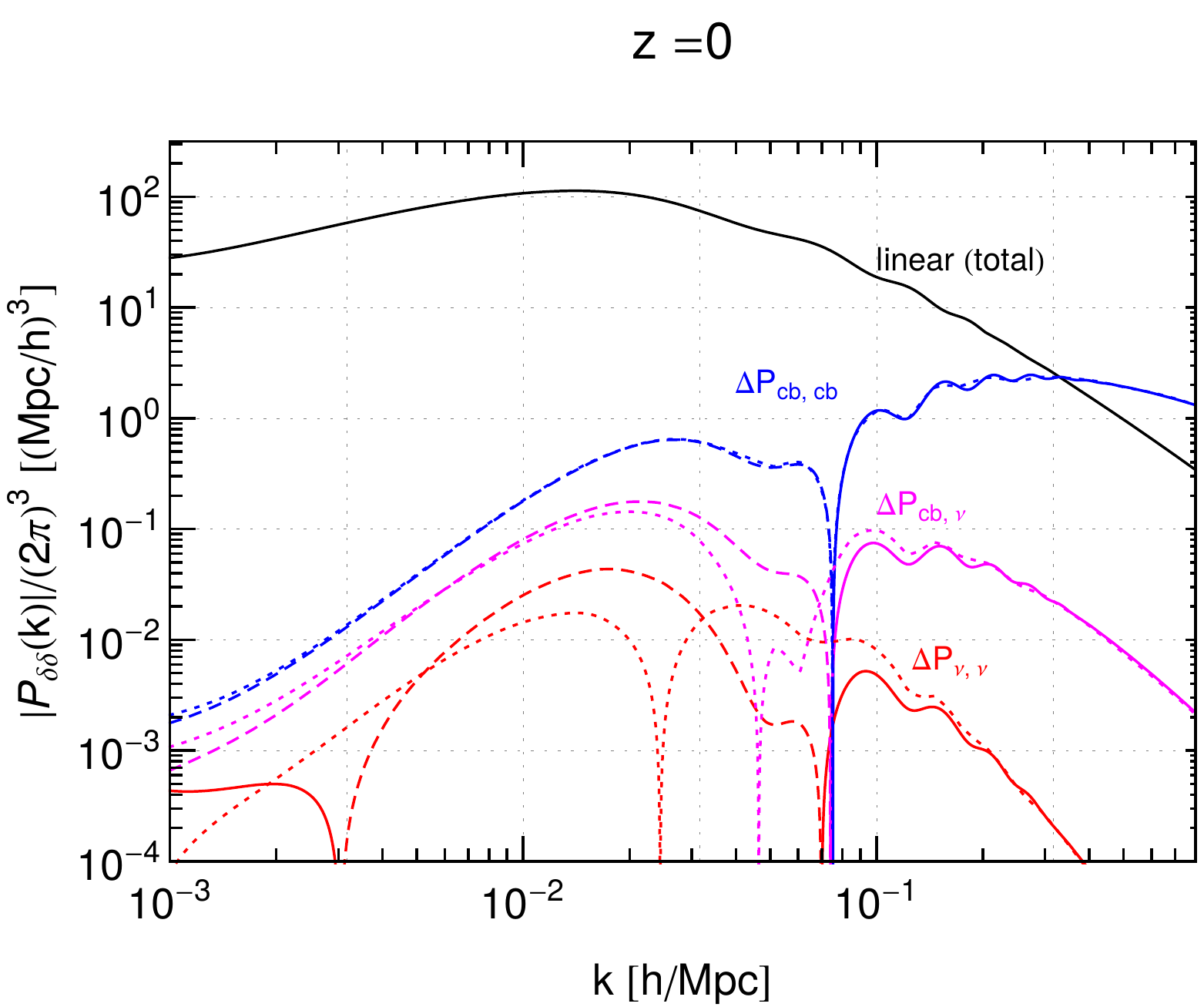}
\includegraphics[width=0.49\textwidth]{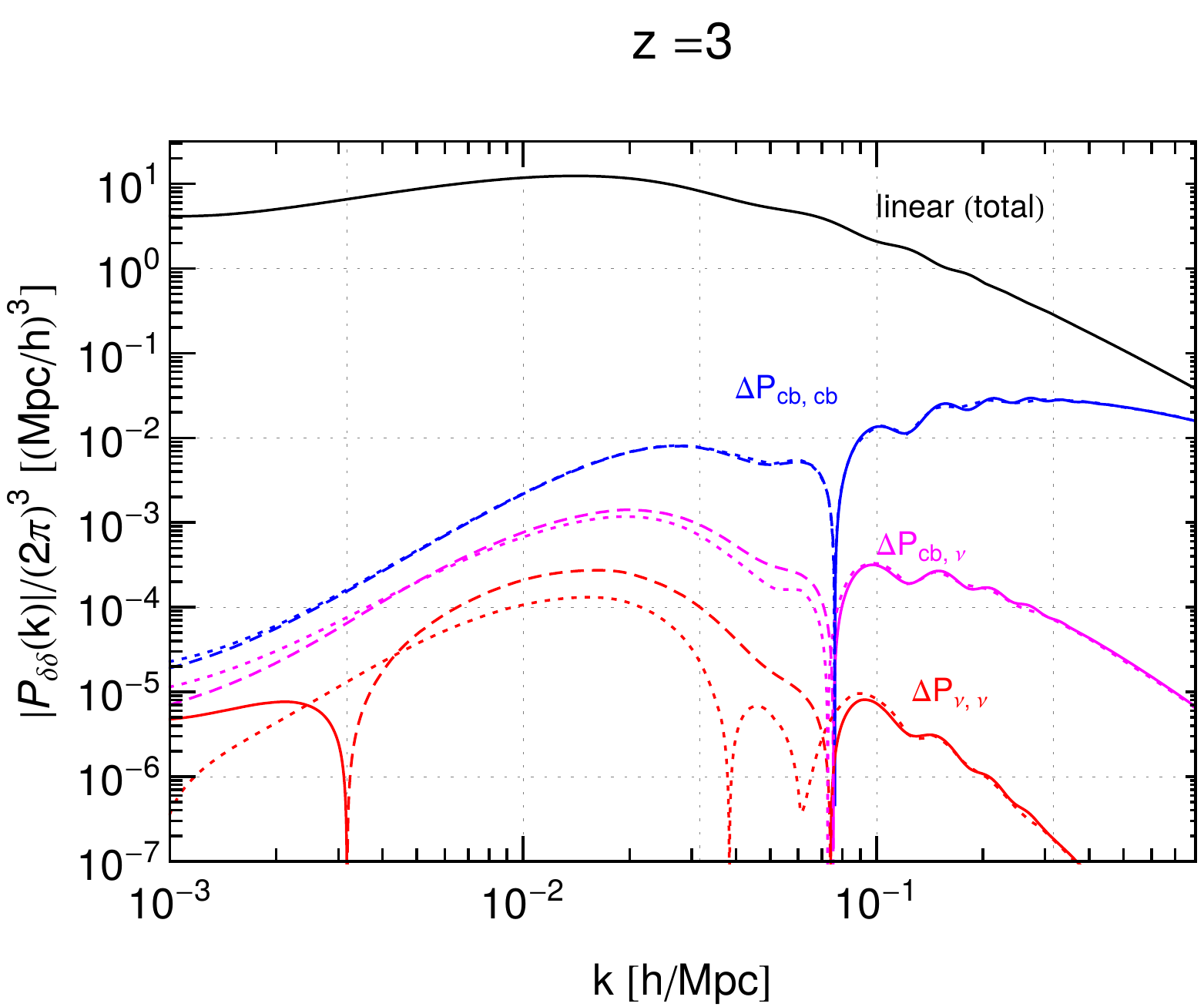}
\caption{\label{fig:nl} Nonlinear corrections to the density power spectra $\Delta P_{ab}=P_{ab}-P_{ab,lin}$ for cold/baryonic matter $P_{cb,cb}$ (blue),
for neutrinos $P_{\nu,\nu}$ (red), and for the cross correlation $P_{cb,\nu}$ (magenta) at redshift $z=0$ (left panel) and
$z=3$ (right panel). The black line shows the linear matter power spectrum. Solid lines correspond to wavenumbers for which $\Delta P>0$, and dashed to $\Delta P<0$. Dotted lines show the corresponding results obtained when taking the full scale dependence for the bispectrum into account.}
\end{figure}

The total matter power spectrum is given by
\begin{equation}
  P(k,z) = (1-f_\nu)^2 P_{cb,cb}(k,z) + 2f_\nu(1-f_\nu) P_{cb,\nu}(k,z) + f_\nu^2 P_{\nu,\nu}(k,z) \;.
\end{equation}
All three contributions can be decomposed into a linear part and a non-linear contribution. The two-fluid description allows us to obtain non-linear corrections to the three (CDM/baryon, neutrino and cross-correlation) spectra. In Fig.~\ref{fig:nl} we show the non-linear corrections obtained from the two-fluid flow equations for these three contributions (dashed/solid lines for negative/positive corrections). 

First of all, we stress that the non-linear corrections are suppressed compared to the linear power spectrum in the limit $k\to 0$ by a relative factor $k^2$ (except for $P_{\nu,\nu}$, see discussion below). This is a consequence of the consistent cancellation of soft effects as discussed above. For this cancellation to occur, it is important to treat all fluid components on an equal footing, and in particular to include non-linear terms involving neutrinos for the computation of $P_{cb,cb}$. 

As expected, the correlators involving neutrinos are suppressed due to free-streaming on small scales $k\gtrsim k_{FS}$. It is interesting to note that the suppression is effective to a certain extent also for $k\lesssim k_{FS}$, because of non-linear mixing of Fourier modes. Together with the small factor $f_\nu$, the non-linear corrections to $P_{cb,\nu}$ and to $P_{\nu,\nu}$ can therefore be safely neglected when one is interested in percent accuracy. 

\begin{figure}
\includegraphics[width=0.49\textwidth]{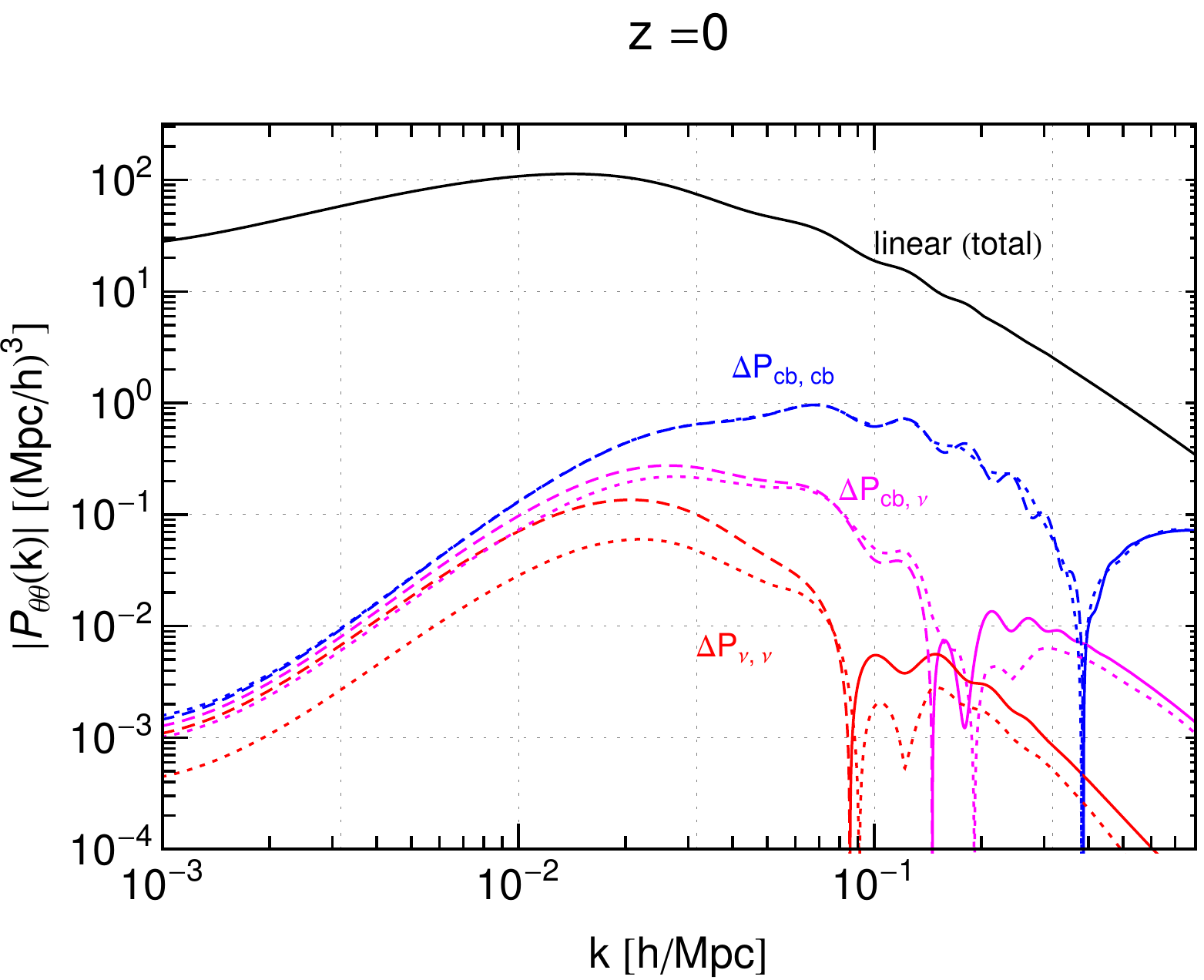}
\includegraphics[width=0.49\textwidth]{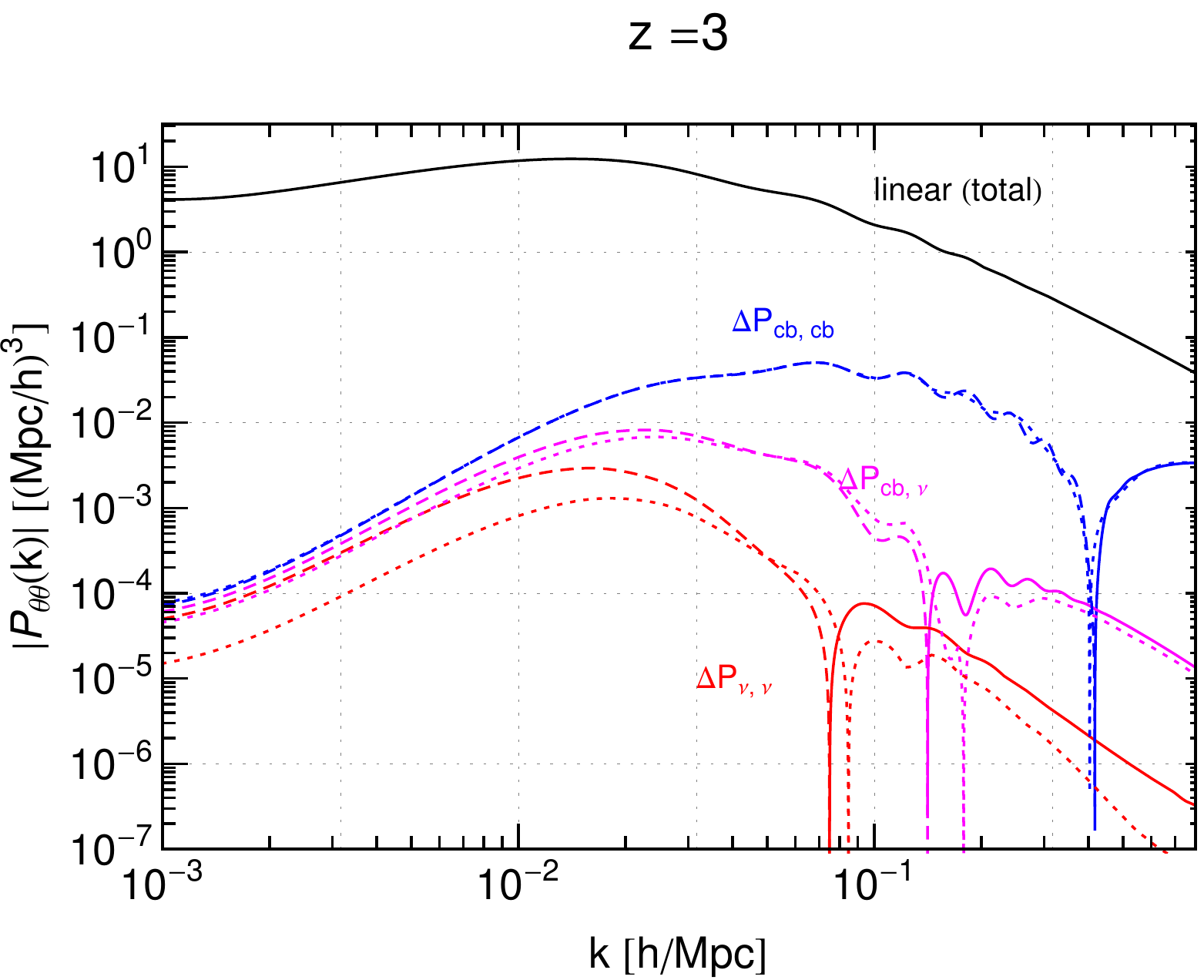}
\caption{\label{fig:nl_theta} As Fig.~\ref{fig:nl}, but for the velocity power spectrum.}
\end{figure}

Fig.~\ref{fig:nl_theta} shows the corresponding correlators for the velocity divergence power spectrum. The behaviour is qualitatively similar.

\subsubsection{Bispectrum with full scale dependence} \label{sec:bispectrum}

As discussed in Sec.~\ref{sec:sol}, the flow equations are based on time evolution equations for the power spectrum $P$ and the bispectrum $B$. Following \cite{Pietroni:2008jx}, the latter are solved very efficiently by neglecting the scale-dependence of the linear propagation matrix $\Omega(k,\eta)$ within the evolution equation for the bispectrum. This simplification has been used in previous studies focussing on the impact of neutrinos \cite{Lesgourgues:2009am,Upadhye:2013ndm}, and we also based our computation within the two-fluid scheme on this approximation. Therefore, one may wonder how large is the error introduced by this simplification. As discussed above, the only source of scale dependence in $\Omega(k,\eta)$ is the free-streaming term for the neutrino component.

\begin{figure}
\begin{center}
\includegraphics[width=0.5\textwidth]{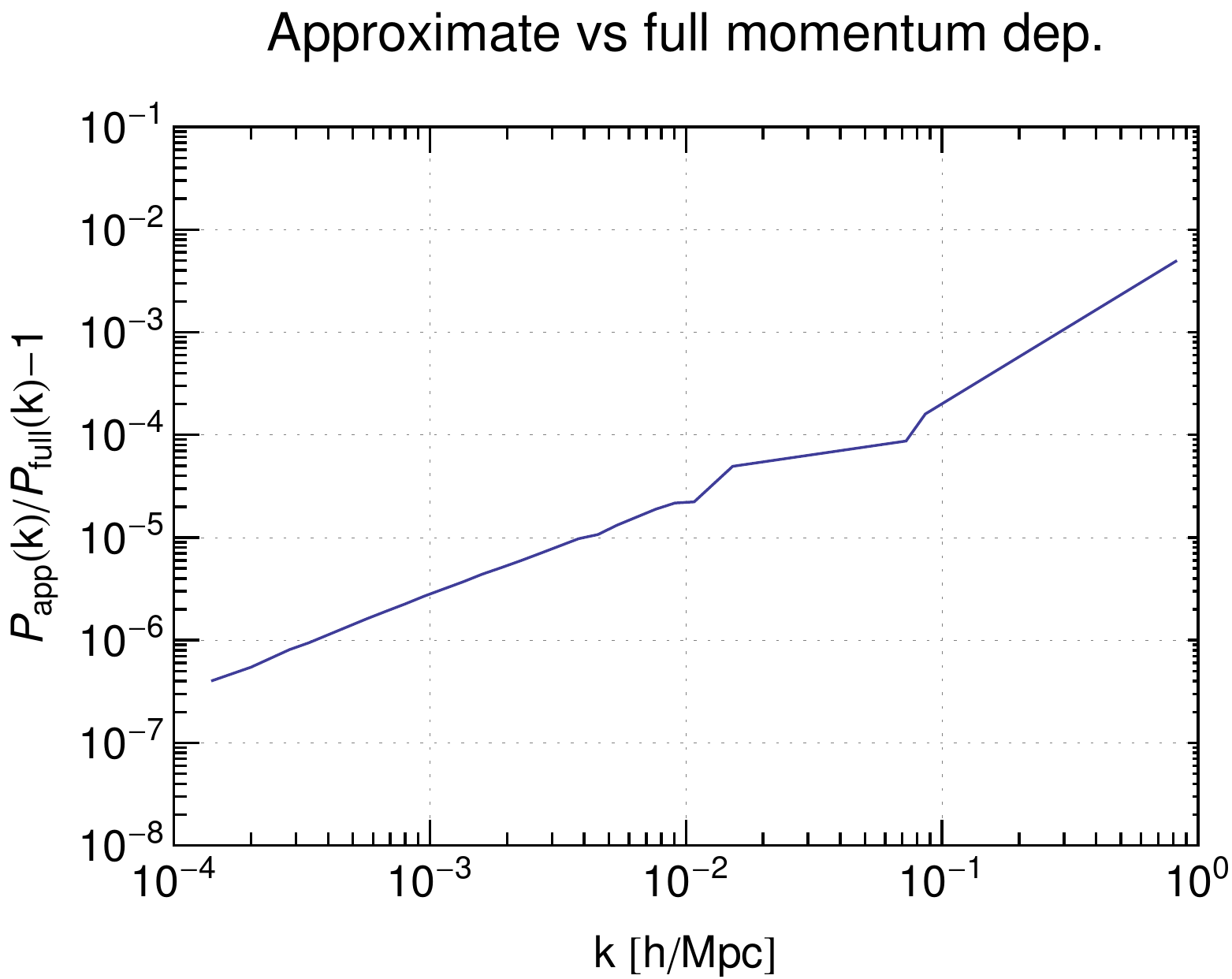}
\end{center}
\caption{\label{fig:PfullMomOverPwithI} Relative deviation of the matter power spectrum ($z=0$) obtained using the full scale-dependent propagator in the evolution equation for the bispectrum (\ref{eq:1loopdynB}) versus the simplified evolution equations (\ref{eq:flow}) introduced in \cite{Pietroni:2008jx}.}
\end{figure}

To test the impact of the above approximation, we also solved the fully scale dependent equation for the bispectrum (\ref{eq:1loopdynB}), and used the result to integrate (\ref{eq:1loopdynP}) for the power spectrum. Since this has to be done for each configuration of the absolute values and the angle between the external wavenumber $k$ and the loop momentum $q$ separately, it is computationally significantly more expensive. The corresponding results for the various power spectra are shown as dotted lines in Figs.~\ref{fig:nl} and \ref{fig:nl_theta}. We observe that there are large deviations for correlators including neutrinos, $P_{cb,\nu}$ and especially $P_{\nu,\nu}$. For the latter, the full equations also yield the expected $k^2$-scaling for $k\to 0$. Therefore, if one is interested in these power spectra, it is important to take the full momentum dependence in the equation for the bispectrum into account. In contrast to this, the CDM/baryon power spectrum $P_{cb,cb}$ is only mildly affected. In Fig.~\ref{fig:PfullMomOverPwithI} we show the relative deviation of the total matter power spectrum (linear+nonlinear correction) compared to the result obtained with the simplified evolution equations (\ref{eq:flow}) at $z=0$. The differences are below the percent level for $k\lesssim 1\,h/$Mpc. Note that the result for the nonlinear contributions (which are suppressed relative to the linear one for small $k$) differ by up to ${\cal O}(10\%(50\%))$
for $k\lesssim 10^{-3}(10^{-4})\,h/$Mpc. Nevertheless, the approximate treatment of the scale dependence seems to be well justified for the matter power spectrum.

\subsubsection{Comparison with approximate schemes}\label{sec:app_vs_full}

For comparison, we compute non-linear corrections for four approximate schemes, that all neglect non-linearities in the neutrino component, and have been discussed in the previous literature.
\begin{enumerate}
\item External source scheme: This scheme takes neutrinos into account only via the contribution of the linear neutrino density perturbations to the gravitational potential. Therefore neutrinos act as an external source for the gravitational force exerted on the CDM/baryon component, in addition to its self-gravity. This approximation scheme violates total momentum conservation. Here we implement it by setting non-linear terms involving neutrinos to zero, $\gamma_{343}=\gamma_{334}=\gamma_{444}=0$, which corresponds to a slight generalization compared to \cite{Wong:2008ws}.

\item Improved external source scheme \cite{Lesgourgues:2009am}: Similar to the previous scheme, but instead of using the \emph{linear} neutrino density in the Poisson equation, its contribution is approximated by
\begin{equation}
 \delta_\nu \to \delta_{cb} \times \frac{\delta_\nu^{lin}}{\delta_{cb}^{lin}} \;, 
\end{equation}
where $\delta_{cb}$ is the full non-linear density contrast. In other words, the \emph{ratio} of neutrino and CDM/baryon density perturbation is approximated by the linear result. This can be described by an effective time- and scale-dependent $\Omega$ matrix for the CDM/baryon component, see \cite{Lesgourgues:2009am}, and can be implemented based on the flow equations for a single fluid. Nevertheless, total momentum conservation is broken also within this scheme. See also \cite{Upadhye:2013ndm} for a recent work using this approximation.

\item SPT-EdS scheme \cite{Saito:2008bp}: A very simple way to estimate non-linear corrections is to ignore completely the different dynamics in the density contrast compared to EdS cosmology, and take into account the effect of neutrinos only via their impact on the linear power spectrum. This amounts to use the standard expressions for the SPT Kernels, which are obtained assuming EdS dynamics, together with the linear spectrum obtained in presence of neutrinos. Since this scheme is agnostic to any changes in the dynamics, it is by definition insensitive to any subtleties arising from backreaction, while completely neglecting the time- and scale-dependent propagation due to neutrino free streaming. On the other hand, its simplicity allows one to compute two-loop contributions or other improvements designed for EdS/$\Lambda$CDM cosmology. This was used recently e.g. to extract information on neutrino masses from BOSS data \cite{Beutler:2014yhv}.

\item Adiabatic scheme \cite{Wong:2008ws}: Similar to the external source scheme, but additionally the time- and $k$-dependence of the growth factor is taken into account
only in an approximate way via modified SPT kernels derived in \cite{Wong:2008ws}. 
\end{enumerate}

\begin{figure}
\includegraphics[width=0.49\textwidth]{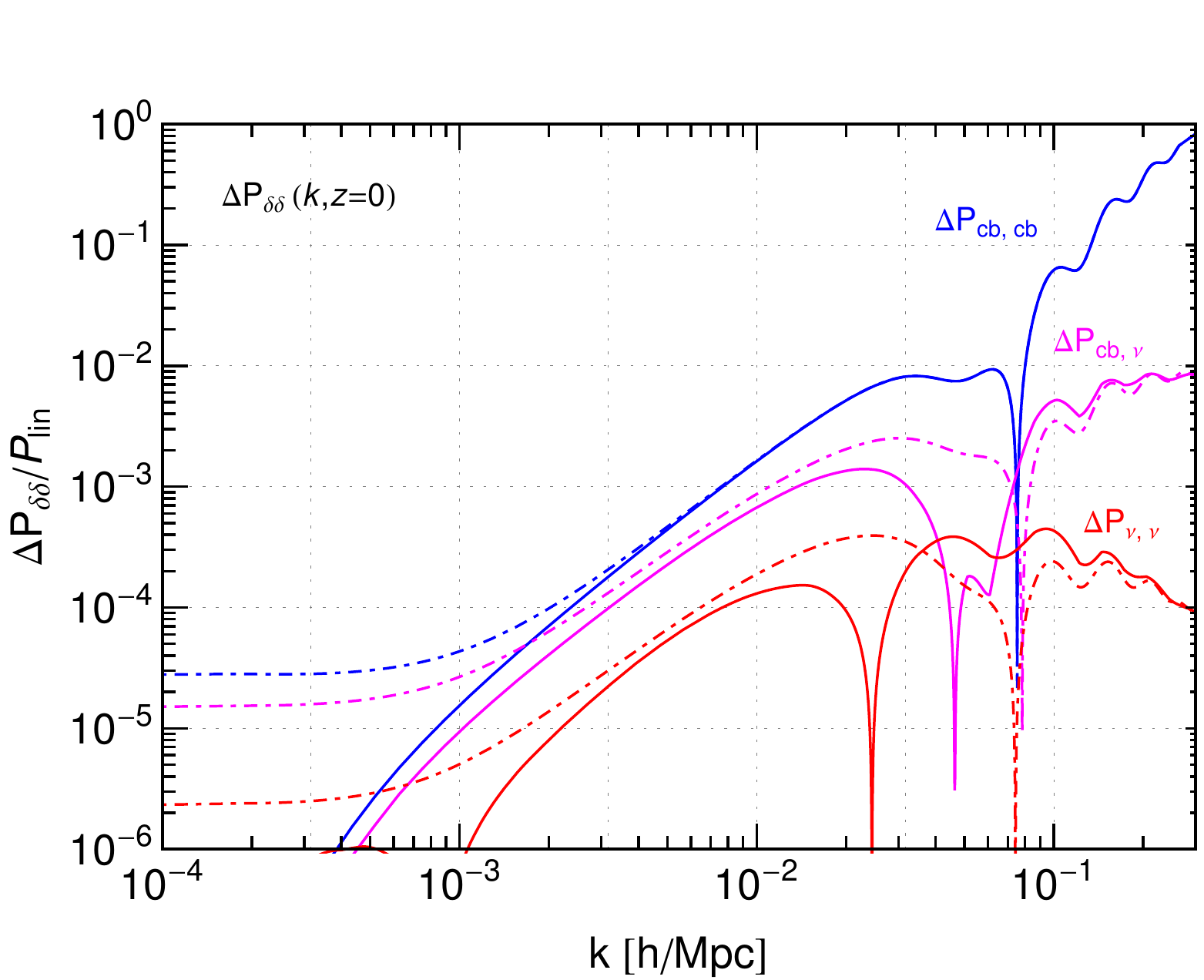}
\includegraphics[width=0.49\textwidth]{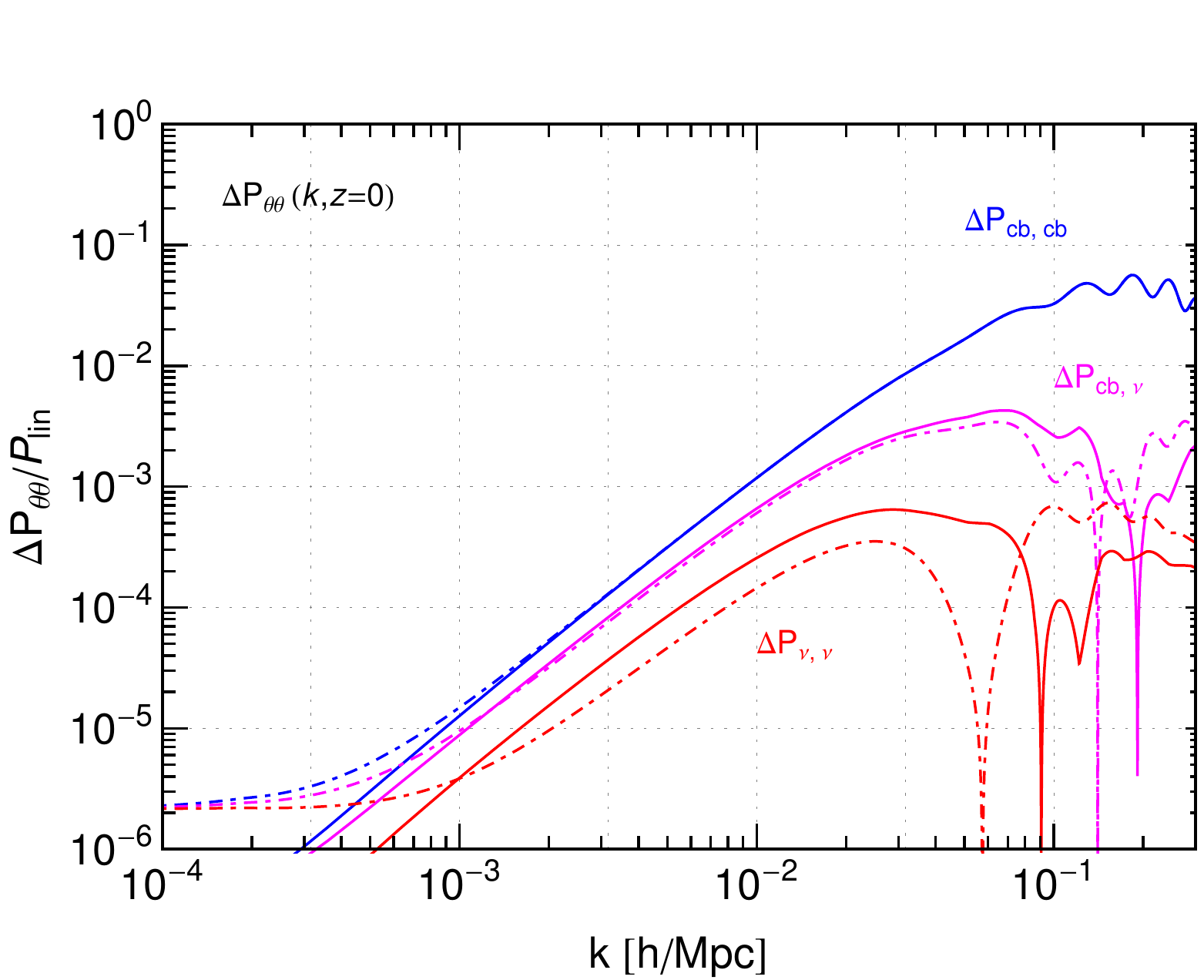}
\caption{\label{fig:app_vs_full1} Comparison of power spectra obtained when treating neutrinos linearly (scheme 1, dot-dashed lines) and for the full two-fluid scheme (solid lines). We show the ratio $\Delta P_{ab}/P_{lin}=(P_{ab}-P_{ab,lin})/P_{lin}$ for CDM/baryonic matter $P_{cb,cb}$ (blue),
for neutrinos $P_{\nu,\nu}$ (red), and for the cross correlation $P_{cb,\nu}$ (magenta) at $z=0$. The left panel corresponds to the density spectra, and the right panel to velocity power spectra. All spectra are normalized by the (total) linear matter power spectrum $P_{lin}$.}
\end{figure}

In Fig.~\ref{fig:app_vs_full1} we compare the CDM/baryon, neutrino and cross power spectra obtained from scheme 1 (dot-dashed lines) with the full two-fluid results (solid lines). The results are normalized to the linear spectrum, such that the $k^2$-scaling for small $k$ is clearly visible for the full solutions. However, the approximate solutions exhibit a qualitatively different behaviour in the large-scale limit. The reason is that neglecting non-linearities in the neutrino field generates a spurious `dipole perturbation', as discussed in Sec.~\ref{sec:dipole}, and spoils the cancellation of soft effects discussed in Sec.\,\ref{sec:IRcancellation}, which together leads to an ${\cal O}(k^0)$ scaling. Additionally, the approximate treatment leads to large deviations in the power spectra involving the neutrinos. Nevertheless, the result for $P_{cb,cb}$ agrees well with the full treatment for $k\gg {\cal O}(10^{-2}-10^{-1})\,h/$Mpc.

\begin{figure}
\begin{center}
  \includegraphics[width=0.75\textwidth]{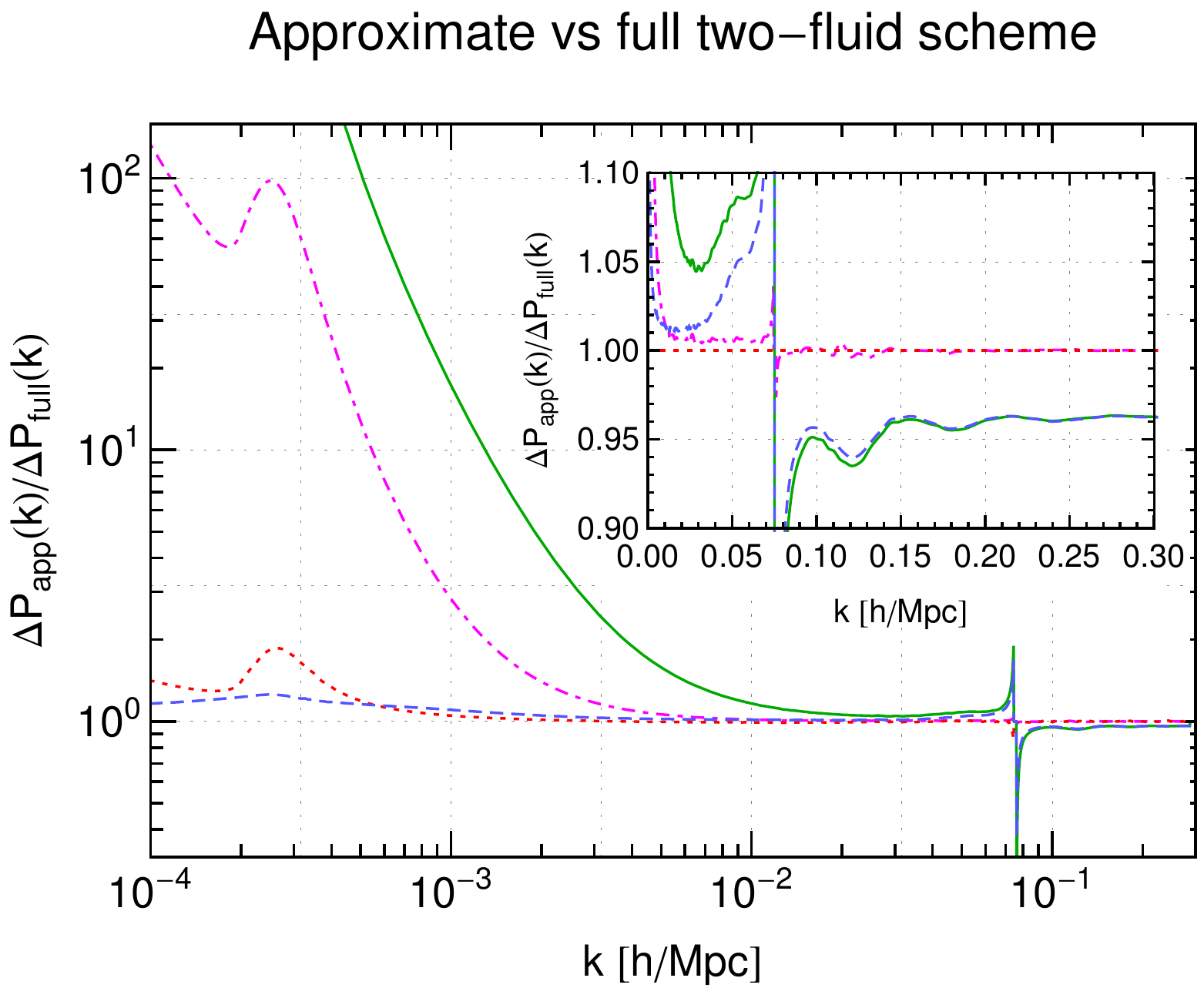}
\end{center}
\caption{\label{fig:app_vs_full} Relative deviation of the non-linear corrections $\Delta P = P - P_\mathrm{linear}$ between various approximate schemes for including massive neutrinos, compared to the full two-fluid scheme ($z=0$). The magenta dot-dashed line corresponds to a linear approximation for the neutrino component as an external source for the matter perturbations (scheme 1). The red dotted line also corresponds to a scheme where the neutrino is treated effectively linearly, but where the \emph{ratio} between neutrino and matter density is assumed to be equal to the corresponding ratio of the linear densities (scheme 2). 
Finally, the blue dashed line corresponds to SPT with EdS kernels, for which the influence of neutrinos is taken into account only via the modification of the linear spectrum that enters the one-loop computation (scheme 3), and the green solid line shows the result obtained based on modified SPT kernels (scheme 4). The inset shows a zoom-in of the right part of the figure. Note that the spike near $k=0.1\,h/$Mpc occurs because the non-linear correction changes sign and goes through zero, which leads to large \emph{relative} deviations.
}
\end{figure}

Let us now discuss also the other approximate schemes 2-4, which in addition to treating neutrinos linearly also reduce the equations to an effective one-fluid form.
Consequently, they can only be used to compute non-linear corrections to $P_{cb,cb}$, while the neutrino and cross correlation are approximated by their linear contribution, $\Delta P_{cb,\nu}\to 0$ and $\Delta P_{\nu,\nu}\to 0$. In Fig.~\ref{fig:app_vs_full} we show the non-linear corrections to the \emph{total} matter power spectrum for all four schemes described above, normalized to the full result. Note that we show the ratio of the \emph{nonlinear corrections} in Fig.~\ref{fig:app_vs_full}, i.e. not the sum of linear and nonlinear part, in order to clearly demonstrate the different scaling for $k\to 0$. Scheme 2 behaves similar to scheme 1 for $k\gg {\cal O}(10^{-2}-10^{-1})\, h/$Mpc, and agrees better with the full result for small $k$. Nevertheless, in principle it can be affected by the spurious `dipole' contributions for $k\to 0$. On the contrary, for scheme 3 one expects a correct scaling limit for $k\to 0$, but quantitative deviations due to the drastic simplifications~\cite{Upadhye:2013ndm}. Indeed, this is confirmed by our numerical results. In particular, the non-linear correction is underestimated by $4-5\%$ in the BAO regime (see inset in Fig.~\ref{fig:app_vs_full}). Finally, scheme 4 yields results that are close to scheme 3 for large $k$, and exhibits spurious behaviour for $k\to 0$.

\begin{figure}
\includegraphics[width=0.49\textwidth]{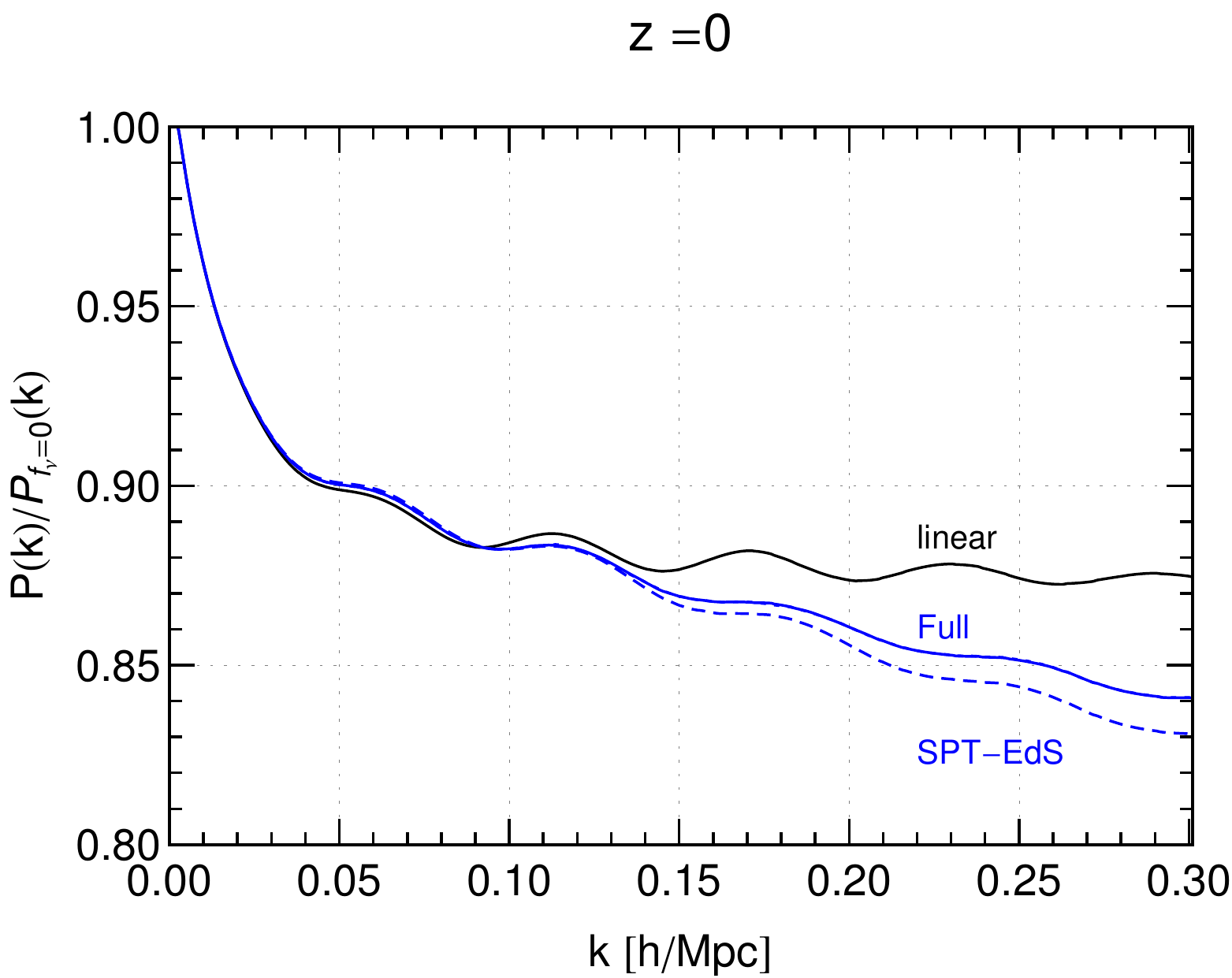}
\includegraphics[width=0.49\textwidth]{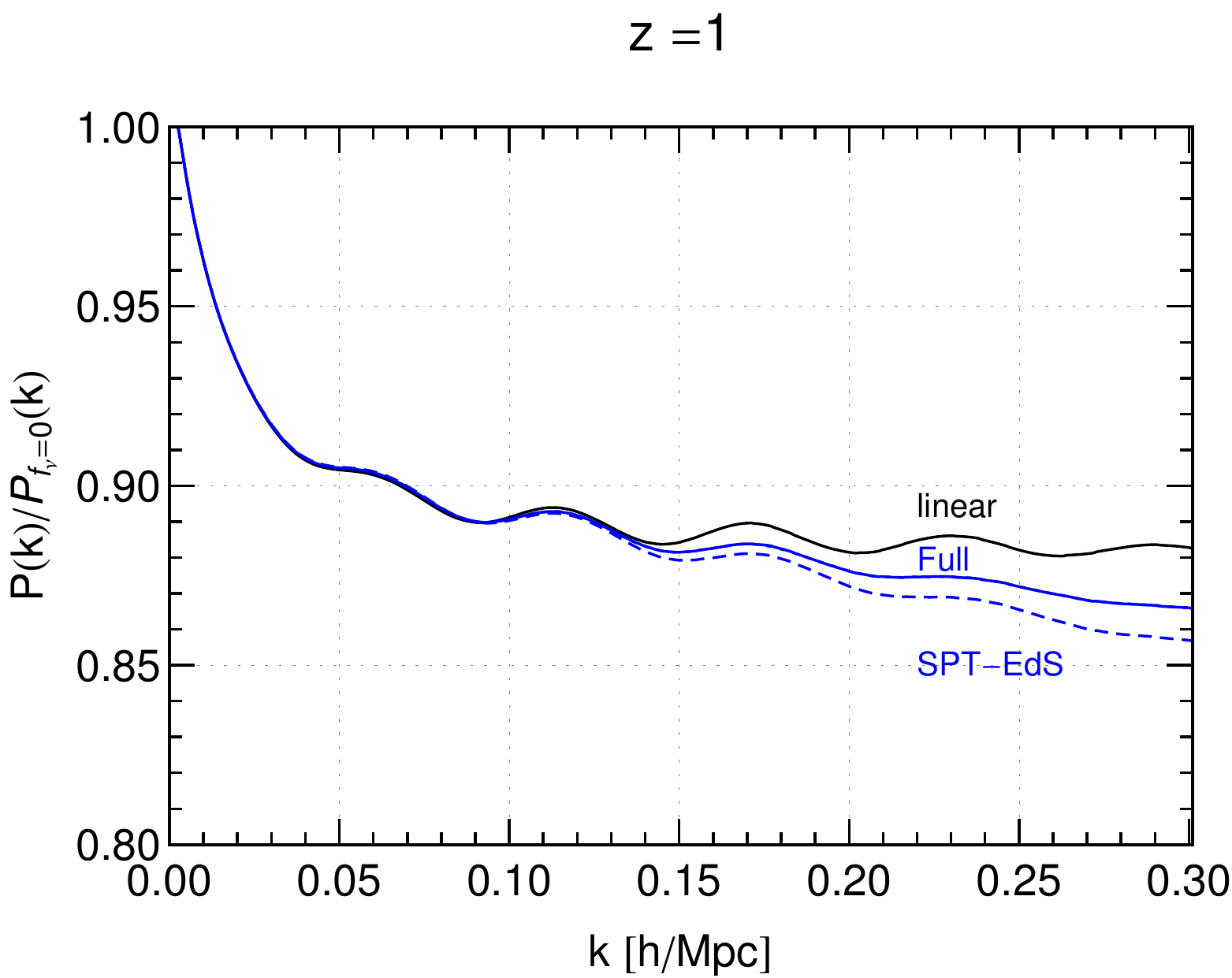}
\caption{\label{fig:ratio} Ratio of the total matter power spectra $P(k,z)/P_{f_\nu=0}(k,z)$ relative to the case with massless neutrinos for $z=0$ (left panel) and $z=1$ (right panel). The blue solid line corresponds to the non-linear solution based on the full two-fluid scheme, and the blue dashed line is the result obtained assuming SPT with EdS kernels (scheme 3). The black line shows the linear power spectra.}
\end{figure}

The ratios of the total matter power spectra (linear+nonlinear correction) for the case with massive neutrinos relative to the massless case are shown in Fig.~\ref{fig:ratio}. Compared to the linear case, the suppression of power on small scales is slightly more pronounced. This behaviour is in general agreement with previous works (see e.g. \cite{Saito:2008bp,Wong:2008ws,Lesgourgues:2009am,Upadhye:2013ndm}), although a quantitative comparison goes beyond the purpose of this work. For comparison, we also show the corresponding result obtained when using scheme 3 (blue dashed line). Note that the result for the ratio when using scheme 1 or 2 agrees reasonably well with the two-fluid solution for $k\gtrsim {\cal O}(10^{-2})\,h/$Mpc, while scheme 4 is very similar to scheme 3 in this range. They are therefore not shown.

Even for the simple SPT-based scheme (scheme 3), the non-linear corrections deviate by more than one percent for all wavenumbers with respect to 
the full result, see Fig.~\ref{fig:app_vs_full} and Fig.~\ref{fig:ratio}. It is thus important to use a scheme taking the time- and scale-dependence of the linear propagator into account if one is interested in percent accuracy. This is achieved by the time-flow equations used for the two-fluid scheme and for the approximate schemes 1 and 2. If one is interested in wavenumbers $k\gtrsim {\cal O}(10^{-2})\,h/$Mpc, it seems safe to use approximation scheme 2. However, it is important to keep in mind that the behaviour for $k\to 0$ is spurious in the approximate schemes 1 and 2, especially when one attempts to use these schemes for computing higher non-linear orders, e.g. by taking the trispectrum into account. Additionally, this is also important to correctly describe the impact of neutrinos in effective approaches or resummation schemes beyond perturbation theory.

\section{Conclusions}\label{sec:5}

In the near future, large-scale structure surveys will allow us to search for the scale-dependent suppression of the matter power spectrum
imprinted by the Standard Model neutrino species, and potentially determine the absolute neutrino mass scale even for a hierarchical
mass spectrum. Apart from
observational and statistical limitations, the sensitivity depends on an accurate understanding and modelling of structure formation
in the transition region between linear and nonlinear scales.

In this work we have computed non-linear corrections to the matter power spectrum, taking the time- and scale-dependent free-streaming
length for neutrinos into account. In particular, we adopted a fluid approach with two components, one representing
pressureless matter (CDM/baryons) and the other neutrino perturbations. To compute the non-linear evolution, we extended the time-flow
framework to take a time- and momentum-dependent effective pressure term for the neutrino component into account, and solve the
evolution equations for the power- and bi-spectrum using the dynamical one-loop approximation. We initialize
the power spectrum using the full linear solution of the Boltzmann hierarchy for the neutrinos at redshift $z_i=25$, for which neutrinos
are well in the non-relativistic regime while non-linearities are still small. We also
generate a suitable initial condition for the bispectrum and check that truncating higher moments of the neutrino distribution for
$z<25$ yields an accurate description of the neutrino density and velocity.

The two-fluid approach allows us to take non-linearities in the neutrino component into account.
Although one might naively expect that
neutrinos can be well approximated by linear theory, we find that this is only partially correct.
In particular, their non-linear mode coupling is important to ensure  judicious cancellations in the perturbative calculation. The latter are important to correctly implement the `decoupling' of small-scale perturbations, which ensure that the non-linear corrections scale as $\Delta P/P_{lin} \propto k^2$ for $k\to 0$
relative to the linear power spectrum. In contrast, when treating neutrinos linearly, the large-scale limit exhibits a spurious behaviour, which
can be traced back to the fact that this simplification is incompatible with basic conservation laws, in particular overall momentum conservation.
As described in Sec.~\ref{sec:dipole}, this spurious behaviour may be real for certain theories beyond $\Lambda$CDM, and we expect to use it to test these models with different observations.

\begin{figure}
\includegraphics[width=0.49\textwidth]{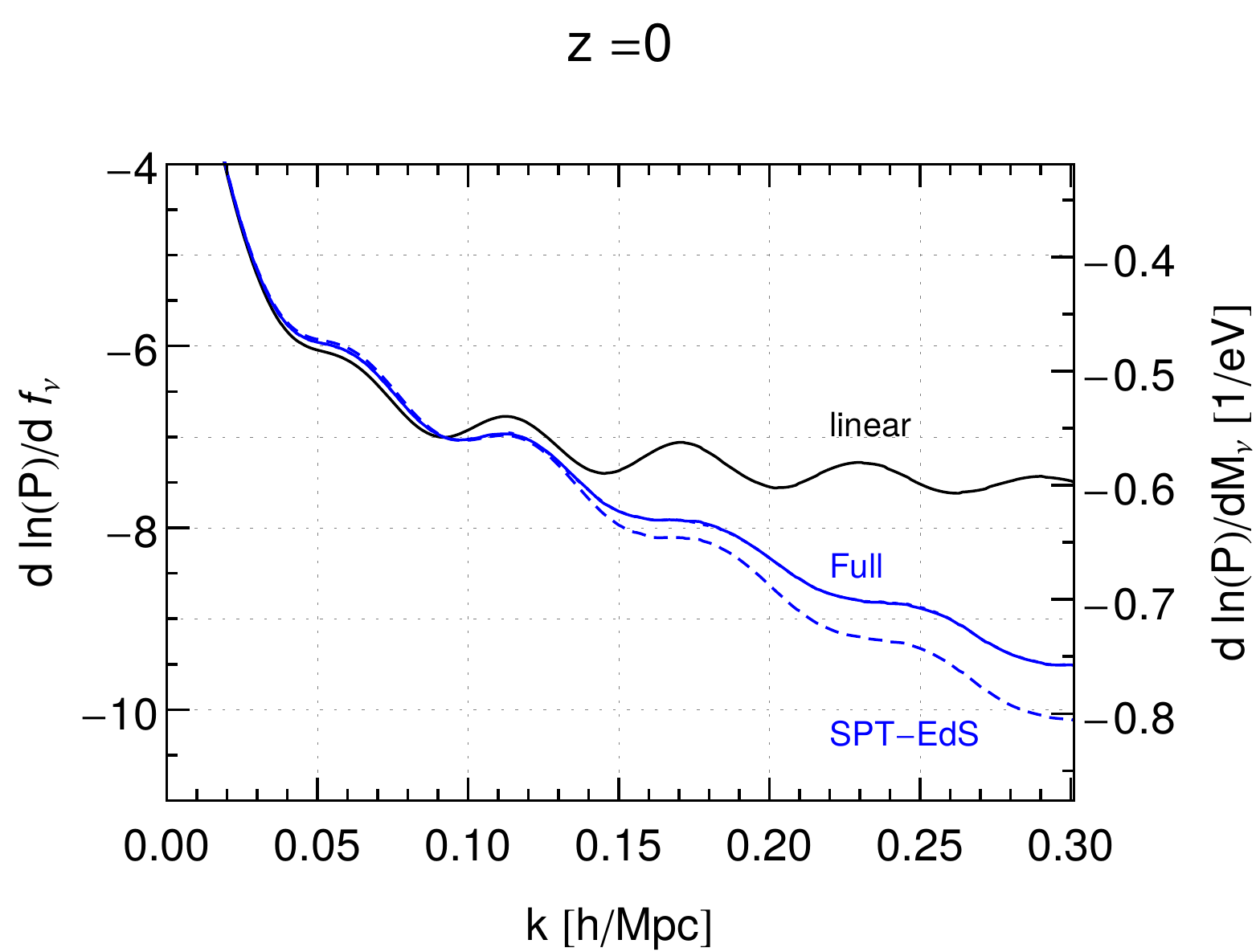}
\includegraphics[width=0.49\textwidth]{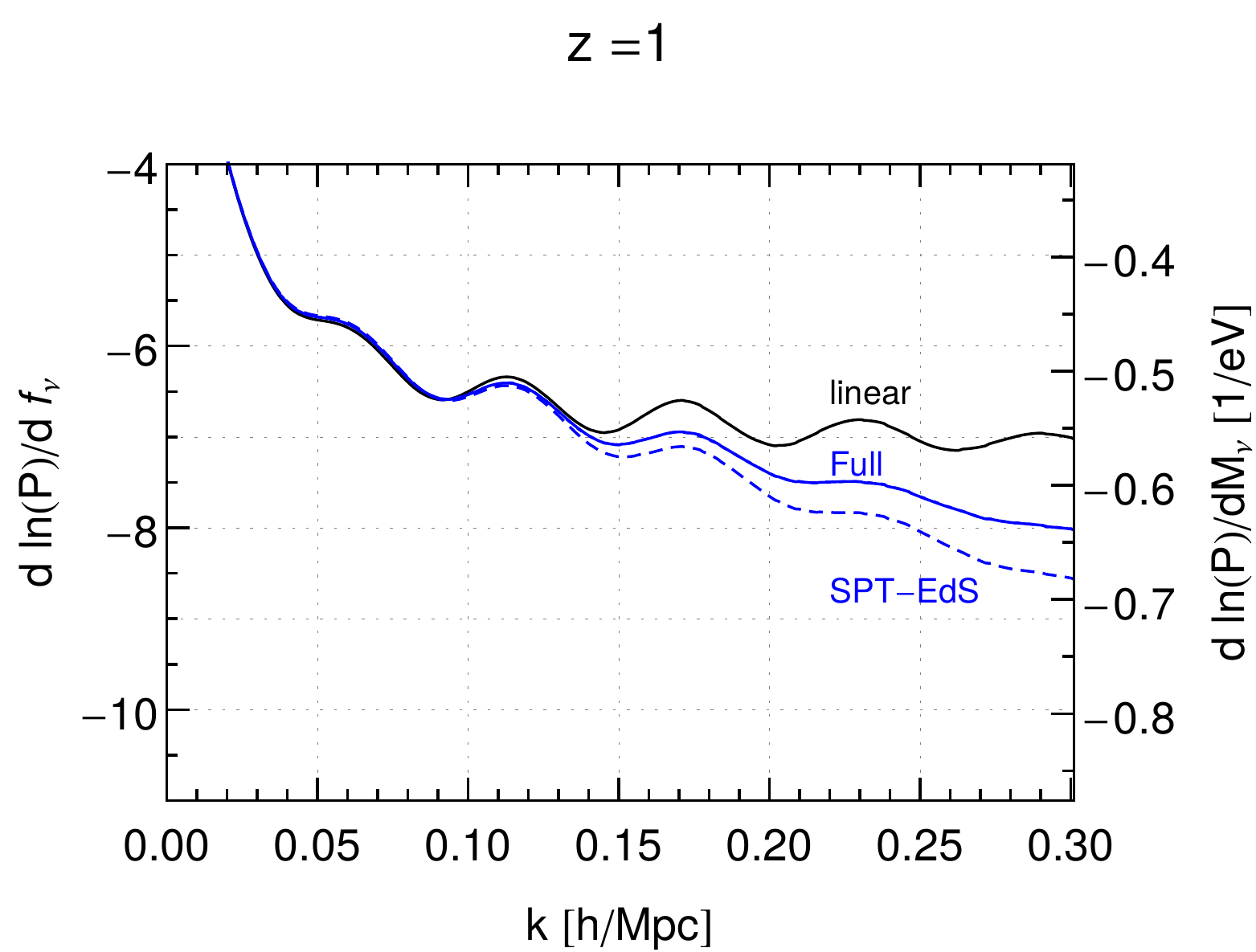}
\caption{\label{fig:dlnPdf} Suppression of the matter power spectrum due to neutrinos, $d\ln(P)/df_\nu$, for $z=0$ (left) and $z=1$ (right). The right axis shows $d\ln(P)/dM_\nu$. The naive expectation for the linear contribution is $d\ln(P)/df_\nu \simeq -8$ at $k\gg k_{FS}$~\cite{Lesgourgues:1519137}.}
\end{figure}

We demonstrate that the two-fluid scheme presented here exhibits the correct scaling behaviour on large scales. Furthermore, we compare our results to various
approximate schemes which treat neutrinos linearly, and find sizeable deviations in the non-linear corrections to the
CDM power spectrum even for a tiny neutrino fraction. For $k \gtrsim {\cal O}(10^{-2}-10^{-1})\,h/$Mpc, the full results deviate from a simple SPT-based scheme, but agree with various approximation schemes that take the time-dependent free-streaming into account. Nevertheless, apart from theoretical consistency, it is worth to mention that the small-$k$ limit is an important input for resummation schemes (e.g. Pad\'e, \cite{Blas:2013aba}) as well as for effective theory approaches (e.g. \cite{Carrasco:2013mua}). Therefore it is crucial to ensure that the approximation exhibits the correct $k^2$-scaling in this regime.
Additionally, we extend the time-flow framework by taking the full scale-dependence of the linear propagator
into account in the equation for the bispectrum, and compute non-linear corrections to the neutrino power spectrum as well as the cross spectrum.

The impact of neutrinos on the matter power spectrum relative to the massless case can be parameterized as 
\be
P(k,z) = P_{f_\nu=0}(k,z) \left[ 1 + \frac{d\ln(P)}{df_\nu}\,f_\nu + \dots \right] \;.
\ee
In Fig.\,\ref{fig:dlnPdf} we compare our result for the coefficient in square brackets with an approximate treatment as well as the linear result.
Equivalently the density fraction $f_\nu$ can be traded for the sum of neutrino masses $M_\nu\simeq 0.21$\,eV\,$\times (f_\nu/0.01673) \times (\Omega_m^0 h^2/0.1348)$.

The correct implementation of the decoupling effect is essential when computing two-loop corrections,
which would be very strongly UV dominated otherwise. The reason is that the $k^2/q^2$-scaling of the integrand kernel (where $q$ is the loop momentum), which is due to intricate cancellations in the perturbative computation, is spoilt when neutrinos are treated linearly. This is also important for any other resummation approach or effective description beyond one-loop. Since the two-fluid scheme discussed in this work possesses a large-scale limit that is consistent with momentum conservation, it is a suitable starting point for going beyond the one-loop order in the future. This will also allow for a meaningful comparison with state-of-the art N-body simulations including neutrinos, e.g. \cite{Villaescusa-Navarro:2013pva}.
Similarly, it would be interesting to compare our results at high 
redshifts  with those of the analytical formulae derived in \cite{Bird:2011rb}. Although such precise computations are needed mainly in view of future large-scale data, which break degeneracies with other parameters and thereby considerably reduce systematic uncertainties, our present results can already be helpful to improve the information on the neutrino mass from current observations \cite{Beutler:2014yhv}.

\section*{Acknowledgements}

We thank R.\,Angulo, B.\,Audren, S.\,Floerchinger, N.\,Tetradis and T.\,Tram for helpful discussions. This work has been partially supported by the German Science Foundation (DFG) within the Collaborative Research Center 676 ``Particles, Strings and the Early Universe''. We thank the Mainz Institute for Theoretical Physics (MITP) and the Benasque Center for Science ``Pedro Pascual'' for their kind hospitality and partial support during the last stages of this work.

\begin{appendix}

\section{Alternative form of evolution equations}\label{app:AltParam}

In this Appendix we describe an alternative formulation of the
continuity and Euler equations (\ref{eq:PTinvectors}) which resembles the one that is commonly
employed within $\Lambda$CDM. Specifically, for a given model featuring
massive neutrinos we consider a corresponding model with massless neutrinos,
but otherwise identical parameters (except for the CDM density which is
adapted in such a way that the \emph{total} density of non-relativistic
species at $z=0$ is identical to the massive model). The growth factor $D_{f_\nu=0}(z)$ of
this model is obtained by solving
\begin{equation}
  ( \partial_\tau^2 +{\cal H}\partial_\tau - \frac32{\cal H}^2\Omega_m(\tau) )
D_{f_\nu=0}(\tau) = 0 \;,
\end{equation}
and is assumed to be normalized to $D_{f_\nu=0}(z=0)=1$. In terms of redshift $z = 1/a-1$,
\begin{equation}
  \left( \partial_\eta^2
+(1+{\cal H}'/{\cal H})\partial_\eta - \frac32\Omega_m \right)
D_{f_\nu=0}(z) = 0 \;,
\end{equation}
where $\eta\equiv\ln(a)$ and ${\cal H}'=d{\cal H}/d\eta = (1-3\Omega_m/2-2\Omega_r){\cal H}$.
Since the total matter density is equal to the case with massive neutrinos, the linear growth
factors agree on large scales for both models. Defining
\begin{equation}
  f \equiv \frac{d\ln D_{f_\nu=0}}{d\ln a}\;,
\end{equation}
and using
\begin{equation}
 \frac{\partial}{\partial\tau} = {\cal H}\,f\,\frac{\partial}{\partial\eta_D} \;,
\end{equation}
one can formally bring the Euler and continuity equations into the same form as
Eq.\,(\ref{eq:PTinvectors}), with $\eta$ replaced by $\eta_D\equiv \ln(D_{f_\nu=0})$ and
a slightly modified definition of the vector $\psi_a$ given by
\begin{equation}
 \psi_1 = \delta_{cb}, \quad \psi_2 = -\theta_{cb}/({\cal H}f), \quad \psi_3 =
\delta_\nu, \quad \psi_4 = -\theta_{\nu}/({\cal H}f) \;,
\end{equation}
and with
\begin{equation}
 \Omega(k,\eta_D) = \left(\begin{array}{cccc}
  0 & -1 & 0 & 0 \\
  -\frac32\frac{\Omega_m}{f^2}(1-f_\nu) & \frac32\frac{\Omega_m}{f^2}-1 &
-\frac32\frac{\Omega_m}{f^2}f_\nu & 0 \\
  0 & 0 & 0 & -1 \\
  -\frac32\frac{\Omega_m}{f^2}(1-f_\nu) & 0 & 
-\frac32\frac{\Omega_m}{f^2}(f_\nu-\frac{k^2}{k_{FS}^2}) &
\frac32\frac{\Omega_m}{f^2}-1 
 \end{array}\right).
\end{equation}
The ratio $\Omega_m/f^2$ has a weak dependence on time, and is often set to
unity (which is exact for EdS). However, since we take a time
dependence in $\Omega(k,\eta)$ into account anyways in order to capture the
time-dependent free-streaming scale, 
we find it more convenient to use the representation (\ref{eq:Omega}).

\end{appendix}

\end{document}